\documentclass[12pt,preprint]{aastex}
\usepackage{graphics}

\begin{document}
\slugcomment{Submitted to ApJ}

\title{Ultra-luminous X-ray Sources in nearby galaxies from ROSAT HRI
observations II. statistical properties}

\author{Ji-Feng Liu, Joel N. Bregman, and Jimmy Irwin}
\affil{Astronomy Department, University of Michigan, MI 48109}

\begin{abstract}

The statistical properties of the ultra-luminous X-ray source (ULX) populations
extracted from the ROSAT HRI survey of X-ray point sources in nearby galaxies
in Paper I are studied to reveal connections between the ULX phenomenon and
survey galaxy properties.
The $logN$--$logS$ relation is used to calculate and remove the influence of
false ULXs from the background and/or foreground objects.
Study of the luminosity functions shows that the regular ULXs below $10^{40}$
erg/sec are an extension of the ordinary HMXB/LMXB population below $10^{39}$
erg/sec in the late-type galaxies, and that the extreme ULXs above $10^{40}$
erg/sec might be  a different population from the regular ULXs.
This survey confirms statistically that the ULX phenomenon is closely connected
to star formation activities, since ULXs preferentially occur in late-type
galaxies rather than in early-type galaxies, and ULXs in late-type galaxies
tend to trace the spiral arms.
Only 5\% of the early-type galaxies host ULXs above $10^{39}$ erg/sec, with
$0.02\pm0.10$ ULX per survey galaxy and $-0.13\pm0.09$ ULXs per
$10^{10}L_\odot$ that are consistent with being zeros.
In contrast, 45\% of the late-type galaxies host at least one ULX, with
$0.72\pm0.11$ ULXs per survey galaxy and $0.84\pm0.13$ ULXs per
$10^{10}L_\odot$.
70\% of the starburst galaxies host at least one ULX, with $0.98\pm0.20$ ULXs
per survey galaxy and $1.5\pm0.29$ ULXs per $10^{10}L_\odot$.
An increasing trend of the occurrence frequencies and ULX rates is revealed for
galaxies with increasing star formation rates.
Two ULX populations, the HMXB-like ULXs as an extension of the ordinary HMXB
population associated with the young stellar population and the LMXB-like ULXs
as an extension of the ordinary LMXB population associated with the old stellar
population, are both required to account for the total ULX population.
It is found that the LMXB-like ULXs dominate the ULX population at low star
formation rates, while HMXB-like ULXs dominate at high star formation rates.
However, an accurate quantitative description of the relative fractions of
HMXB-like ULXs and LMXB-like ULXs is impossible in the current HRI survey due
to the source blending effects, and a Chandra survey of the X-ray point sources
in nearby galaxies would be required for this purpose.

\end{abstract}


\section{INTRODUCTION}

Ultra-luminous X-ray sources (ULXs) are extra-nuclear sources with luminosities
in the range of $ 10^{39}-10^{41}$ erg/sec in other galaxies, and have been
observed by the Einstein Observatory (e.g., Fabbiano et al. 1989), ROSAT (e.g.,
Colbert \& Mushotzky 1999), ASCA (e.g., Makishima et al. 2000), and recently by
the Chandra and XMM-Newton X-ray Observatories in many galaxies (e.g., Kilgard
et al. 2002; Swartz et al. 2004).
As compared to the cases of the X-ray binaries in our Galaxy, which are powered
by accretion onto neutron stars or stellar mass black holes and have
luminosities of $ 10^{33}-10^{39}$ erg/sec, the luminosities of ULXs require
accreting black holes of masses  $10^2$ -- $10^4$ $M_\odot$ if they emit at
0.1--0.01 of the Eddington luminosity, typical of Galactic X-ray binaries
(Colbert \& Mushotzky 1999).
Such intermediate mass black holes, if they exist, bridge the gap between
stellar mass black holes and supermassive black holes of $10^6$ -- $10^9
M_\odot$ in the center of galaxies.
Alternatively, ULXs could be stellar mass black holes or neutron stars whose
apparent super-Eddington luminosities are due to some special processes. One
suggestion is that radiation pressure-dominated accretion disks with
photon-bubble instabilities are  able to emit truly super-Eddington
luminosities (Begelman 2002). Another suggestion is that beaming effects can
produce the observed luminosities of ULXs (King et al. 2001).

The leading goals in ULX studies are to determine the masses of the primary, to
understand how and where they form, and to find out how they emit at such high
luminosities.
In the last few years observations of many ULXs in nearby galaxies have been
made to address these questions, and important clues have been revealed on
their spectra (e.g., Makishima et al, 2000; Miller et al. 2003), their
preferential occurrence in galaxies with current star forming activities (e.g.,
Kilgard et al. 2002; Zezas et al. 2002; Gao et al. 2003), their short-term and
long-term variabilities, periodic variations (e.g., Sugiho et al. 2001; Bauer
et al.  2001; Liu et al. 2002), and quasi-periodic oscillations (e.g.,
Strohmayer \& Mushotzky 2003; Soria et al. 2004; Soria \& Motch 2004; Liu et
al. 2004).
Aside from in-depth studies of some well-known ULXs, the study of a complete
sample of ULXs, with incompleteness carefully calculated, is necessary to study
the defining properties of ULXs as a class, and  to calculate the occurrence
rates and luminosity functions that will place constraints on models on how
ULXs form and evolve.
%


With observations of individual ULXs in nearby galaxies accumulating, now it is
becoming feasible to construct samples with statistically significant number of
ULXs.
So far there are only a few such studies. 
Colbert \& Ptak (2002, hereafter CP2002) analyzed the HRI observations to
search for ULXs in a sample of 9999 galaxies in the Third Reference Catalog of
galaxies (RC3; de Vaucouleurs et al.  1991) with $cz<5000$ km/sec.  
They found 87 ULXs in 54 galaxies, with 37 in early-type galaxies.
Based on this catalog, Ptak \& Colbert (2004) found that $\sim$ 12\% of
galaxies contain at least one ULX with $L_X \ge 10^{39}$ erg/sec and $\sim$ 1\%
of galaxies contain at least one ULX with $L_X \ge 10^{40}$ erg/sec in 2--10
keV. 
However, many ULXs in the CP2002 catalog are projected far from the host
galaxies, and may be foreground stars or background AGN/QSOs instead (Irwin et
al. 2004). 
Recently Swartz et al. (2004) analyzed the archive Chandra
ACIS observations for 82 nearby galaxies, in which they found 154 extranuclear
point sources with $L_X \ge 10^{39}$ erg/sec in 0.5--8 keV as ULX candidates.
They estimate a ULX rate of 
$1.5\pm0.55$ ULXs per $10^{10} M_\odot$ 
for spirals, and $0.55\pm0.10$ ULXs per $10^{10} M_\odot$
for early type galaxies.  
Based on a rough comparison of the cumulative luminosity functions of the ULXs
to the Chandra Deep Field results they conclude $\sim$ 14\% (44\%) of the ULXs
in spiral (early-type) galaxies may be background objects.


We have carried out a survey of ULXs in nearby galaxies with ROSAT HRI archival
data (Liu \& Bregman, 2005; hereafter Paper I), which takes advantage of the
moderate spatial resolution of HRI (with on-axis FWHM of $<5^{\prime\prime}$)
and the large sky coverage ($\sim$2\%) of the data archive.
This survey encompasses 313 galaxies within 40 Mpc with isophotal diameters
$>1^\prime$ observed by 467 HRI observations, with an average of 2.2
observations per galaxy.
Uniform data reduction procedures were applied to all observations to detect
point sources with a wavelet detection algorithm, for which simulations were
run to understand its behaviors on HRI observations, including the false
detection rates, the detection thresholds, and the correction factors to
correct detected counts to true source counts based on the detection
significance and the off-axis angles of detected sources.
As a result, 106 extra-nuclear point sources with $L_X \ge 10^{39}$ erg/sec in
0.3--8 keV defined as ULX candidates are found in the $D_{25}$ isophotes of 63
galaxies, with ten already identified as QSOs or stars. There are 110 ULX
candidates found between 1--2 $\times D_{25}$ of 64 galaxies, with 16 already
identified as QSOs or stars.

To minimize the contamination from background and/or foreground objects, in
Paper I we constructed a clean sample of 109 ULXs from the ULX candidates by
selecting only those within the $D_{25}$ isophotes or with apparent connections
with the host galaxy, and excluding those identified as QSOs and stars. 
The clean sample forms a good basis for studies on ULX's environments and
identifications, since the contamination is small due to exclusion of known
QSOs and stars.
The trend that ULXs are preferentially found in spiral galaxies is clearly
shown in the clean sample, with 49 out of 181 spiral galaxies in our survey
host 89 ULXs, while 4 out of 93 early-type galaxies host 7 ULXs if we exclude
the 8 ULXs in two peculiar lenticular galaxies NGC1316 and NGC5128.
However, it is inappropriate to compute the occurrence frequency of ULXs in
spiral (early-type) galaxies with this clean sample, because the selection
criteria for ULXs in this sample are not uniform, and the contamination from
background and foreground objects, though small, is uncertain.

In this paper, we compute the occurrence rates and the luminosity functions of
ULXs in different types of galaxies with the contamination carefully calculated
and subtracted.
In Section 2 we present the methodology used in our analysis, including how ULX
samples are constructed from the HRI survey, how the contamination is
calculated, what aspects of the samples are studied, and how the ULX rates are
normalized by the surveyed blue light.
We compare in Section 3 the statistical properties of ULX samples in different
types of galaxies, including early-type galaxies, late type galaxies, starburst
and non-starburst galaxies, and galaxies with different star formation rates.
We study the radial distribution of ULXs with respect to the isophotal ellipses
of the galaxies in Section 4.
The paper is concluded by discussions on  our results in Section 5.


\section{METHODOLOGY}

Statistical analysis of large samples of ULXs is essential to study ULXs as a
class. There have been a few such studies, e.g., CP2002, Kilgard et al. (2002),
Ptak \& Colbert (2004), and Swartz et al. (2004), most of which do not have a
thorough analysis of the contamination from background and foreground objects.
Here we present our method for analyzing the ULX samples from our ROSAT HRI
survey, with particular care on the contamination problem.

\subsection{Construction of ULX samples}

The total HRI survey includes 313 galaxies of different types with different 
properties. 
In our study, we divide the 313 galaxies into groups to study the connection
between ULXs and the properties of the galaxies.
Galaxies are grouped based on the morphological types of the galaxies to reveal
possible trends of ULXs along the Hubble sequence.
To study the connection between the star formation activities and the ULX
phenomenon, we group the galaxies based on whether they are starburst or HII
galaxies, and also based on the star formation rate inferred from the FIR
luminosity.

A complete ULX sample is extracted from a snapshot survey of galaxies, in which
a galaxy is observed in {\it one} observation, by selecting {\it all}
extranuclear X-ray point sources within (parts of) the galaxies with $L_X >
10^{39}$ erg/sec.
Such a sample is an instantaneous sample, which is different from the
collection of ULX candidates from all 313 survey galaxies listed in Table 3 of
Paper I, because the latter is collected from {\it multiple} observations of
galaxies.
Such a sample is also different from the clean sample of ULXs defined in Paper
I, because some extranuclear sources identified as background QSOs and
foreground stars are excluded from the clean sample.

To make a snapshot survey of the galaxies, one observation needs to be selected
for each galaxy.
One way is to select the observation with the longest exposure for each galaxy,
which we refer as the deep survey.
Another way is to select a random observation from all available for each
galaxy to form a random survey.
In our analysis, we compute 20 random surveys, the average of which is
calculated and referred as the average survey.
The ULX samples drawn from the average survey and the deep survey are usually
different, because the X-ray sources vary in luminosity between observations,
and the deep survey has larger low luminosity coverage than the average survey.



The ULX candidates are grouped by their proximity to the galactic center into
those within the $D_{25}$ isophotes, and those  between 1--2 $\times D_{25}$.
In our discussion, we focus on the statistical properties of the the ULX
samples within the $D_{25}$ isophotes of the galaxies, which presumably have a
higher fraction of ``true'' ULXs than the ULX samples between 1--2 $\times
D_{25}$.
To reveal how ULXs are distributed radially in galaxies, we study the ULX
samples within annuli of the isophotes in Section 4, and compare the ULX sample
between  1--2 $\times D_{25}$ to the ULX sample within  the $D_{25}$ isophotes.

In the following sub-sections, we describe how we calculate the number of
contaminating sources based on the $logN$--$logS$ relation, the survey blue
light curve, the ULX rates,  and luminosity functions for the ULX samples.

\subsection{Contamination of ULX samples}

ULXs are usually defined by associating an X-ray source to a galaxy based on
its projected proximity to the galaxy. Such a definition inevitably introduces
into a ULX sample the contamination from foreground and/or background objects,
which must be excluded when studying the statistical properties of the ULX
sample.
A direct way is to make deep optical and/or spectroscopic observations of every
ULX candidate in a sample, identify those foreground and/or background objects,
and exclude them from the sample. 
However, it is impractical to make such studies for all ULXs in a large sample
considering the huge amount of telescope time needed, although optical and
spectroscopic studies of a number of ULXs are essential and have to be made to
determine the nature of the ULX systems.

A practical way to estimate the contamination from background and foreground
objects, as used in this paper, is to make predictions with a logN--logS
relation, and exclude the predicted numbers from the ULX sample in the
statistical studies that follow.
For the HRI observations in this survey, it is appropriate to use the logN-logS
relation derived from ROSAT observations (Hasinger et al. 1998), where the
differential form is $dN/dS = N_1 S^{-\beta_1}$ for $S>S_b$ and $dN/dS = N_2
S^{-\beta_2}$ for $S<S_b$, with $S$ in unit of $10^{-14}$ erg/sec/cm$^2$,
$S_b=2.66\pm0.66$, $N_2 = 111\pm10$, $\beta_1 = 2.72\pm0.27$, $\beta_2 =
1.94\pm0.19$, and $N_1 = 238.1$. 
With this differential relation, the number of contaminating sources in a flux
interval can be calculated with the survey area curve $A($$>$$S)$ that gives
the area (i.e., solid angle) in which sources brighter than the limiting flux
$S$ can be detected in the survey. A typical error of 10\% is associated with
the predicted number of contaminating sources.


The survey area curve $A($$>$$S)$ can be calculated for an observation given
the sensitivity (detection threshold) of the observation.
For our data reduction procedures, the $3\sigma$ detection threshold can be
computed with the background level and the source size that has been derived
from simulations in Paper I as a function of the off-axis angle $\theta$.
A sensitivity map is constructed for each observation by computing the
detection thresholds (in unit of count/sec) at the location of every pixel,
with the average background level computed from the part of the detector with
off-axis angle $\theta<17^\prime$ excluding the detected sources.
The survey area curve $A($$>$$S)$ for a galaxy in this observation can be
computed by summing up the area of the pixels within this galaxy for which the
detection thresholds correspond to flux less than $S$.
Similarly, the survey area curves for parts of a galaxy, e.g., the region
between 1--2 $\times D_{25}$, can be computed by considering only pixels within
those parts.

The detection threshold in count rate is converted to flux $S$ by assuming a
power law spectrum, using Galactic HI column density toward the galaxy, with a
photon index of 2 within 0.5--2.4 keV to be consistent with what was used in
the derivation of the logN--logS relation (Hasinger et al. 1998).
Note that the flux $S$ for use in the logN--logS relation is different from the
apparent flux $F$, used to calculate the luminosity $L$ in Paper I, because for
the flux $F$  we used a power-law spectrum with a photon index of 1.7 within
0.3--8.0 keV to be comparable to Chandra observations.
The flux $F$ is higher than $S$ by a factor of $\sim$3 for the same count
rate, with the factor varying slightly with the Galactic HI column density used
in the conversions.


Given the survey area curve $A($$>$$S)$ for a galaxy in an observation, the
number of background and foreground objects can be calculated by integrating
the product of $dN/dS$ and $A($$>$$S)$ over flux intervals.  The number $
N_b(S_1,S_2)$ over a flux interval of $(S_1,S_2)$ is 
$$N_b(S_1,S_2) = \int_{S_1}^{S_2} {dN \over dS} A(>\!S) dS$$
To estimate the number $N_b(L_1,L_2)$ for a galaxy in an apparent luminosity
bin $(L_1,L_2)$, the luminosity $L$ is converted to $S$ using a function
$S_g(L)$ with the distance to the galaxy, and with proper consideration of the
different conversion factors from the count rate to flux $S$ and flux $F$.
For a group of galaxies, the total number $N_b(L_1,L_2)$ can be
calculated by summing up the numbers for individual galaxy in their respective
observation, i.e., 
$$N_b(L_1,L_2) = \sum_g \int_{S_g(L_1)}^{S_g(L_2)} {dN \over dS} A_g(>\!S) dS$$
Here $S_g(L)$ has a dependence on the distance of the galaxy and the Galactic
HI column density toward the galaxy.

\subsection{Blue light surveyed in ULX surveys}

The occurrence rates of ULXs in different types of galaxies place
constraints on models on how ULXs form and evolve.
The occurrence rate can be directly calculated as the number of ULXs per
galaxy, or calculated as the number of galaxies per unit stellar mass to
account for the variations in the size and stellar mass of galaxies among and
within different types.
The stellar mass content of a galaxy can be inferred from the total light of
the galaxy with the stellar mass-to-light ratio. 
While the mass-to-light ratios have larger variations in the optical than in
the near-infrared (Bell \& De Jong, 2001), in our analysis we use the blue
light from RC3 and calculate the number of ULXs per $10^{10} L_\odot$ of blue light,
due to the lack of a uniform compilation of near-infrared magnitudes for
the survey galaxies.

The survey blue light curve $\pounds_g(L)$ for a galaxy gives the blue
light of the stellar contents in which X-ray sources above $L$ can be detected
in an observation.
To compute the detection threshold in $L$, we calculate the detection threshold
in count rate for each pixel of the HRI image as described in Section 2.2,
convert it to flux $F$ by using a power-law spectrum with a photon index of 1.7
in 0.3--8 keV, and calculate $L = 4\pi D^2 F$ with $D$ as the distance to the
galaxy.
To compute the blue light, we compute the light profile for the galaxy from the
total blue light $\pounds_B$ and the effective radius $R_e$ that encloses 50\%
of the total light.
For two thirds of the survey galaxies, the effective radii are taken from RC3,
with an average $R_e$ of $0.15 \times D_{25}$.  For one third of the survey
galaxies without $R_e$ from RC3, we assume $R_e = 0.15 \times D_{25}$.
The blue magnitudes of 297 survey galaxies are taken from the RC3 catalog, and
converted to blue light $\pounds_B$ in unit of $L_\odot$ with $M_{B\odot} =
5.46$ mag.

The light profile for early-type galaxies is usually expressed in the De
Vaucouleurs $R^{1/4}$ law, which reads $I(R) = I_e
10^{3.33-3.33(R/R_e)^{1/4}}$. Here $I_e$ is the surface brightness at the
effective radius $R_e$, and the total light can be expressed as $7.22\pi R_e^2
I_e$.
With $R_e$ and $\pounds_B$ known, the light profile can be expressed as $I(R) =
{\pounds_B \over 7.22\pi R_e^2} 10^{3.33-3.33(R/R_e)^{1/4}}$.

The light profile for late-type galaxies is decomposed into a De Vaucouleurs
bulge in the form of $I(R) = I_e 10^{3.33-3.33(R/r_e)^{1/4}}$,  and an
exponential disk in the form of $I(R) = I_0 e^{-R/h}$, with $h$ as the scale
height, $I_0$ as the surface-brightness at the galactic center, and the total
light from the disk as $2\pi h^2 I_0$.
The relative prominence of the two components change with the Hubble type T,
and the bulge-to-disk ($B/D$) ratios, taken from Graham (2001) and interpolated
when necessary, change from 25\% at T=0 to 1\% at T=10 with a general decrease
toward later galaxy types.
In spite of the variation in bulge-to-disk ratios, the $r_e/h$ ratio is found
to be quite constant, and we take $r_e/h=0.2$ following Graham (2001).
By assuming the effective radius $R_e$ derived from RC3 encloses 50\% of the
disk light, we find $h = R_e/1.7$, $r_e = R_e/8.5$.  This assumption is
reasonable for most of the survey galaxies for which $B/D \ll 1$; for galaxies
with non-negligible bulges, this slightly over-estimates $r_e$ and $h$.
Given $\pounds_B$ from RC3, we find $I_0 = {\pounds_B \over 1+B/D} {1.7^2 \over
2\pi R_e^2}$, and $I_e = {\pounds_B \over D/B+1 } {8.5^2 \over 7.22\pi R_e^2}$.
With $r_e$,$h$,$I_e$, and $I_0$ calculated from $R_e$ and $\pounds_B$, the
light profile can be calculated with $I(R) = I_e 10^{3.33-3.33(R/r_e)^{1/4}} +
I_0 e^{-R/h}$.

The survey blue light curve $\pounds_g(L)$ for a galaxy can be computed by
summing up the blue light in pixels within the galaxy for which the detection
thresholds correspond to luminosities below $L$.
Similarly, the curve $\pounds_g(L)$ for parts of a galaxy, e.g., the
region between 1--2 $\times D_{25}$, can be computed by considering only pixels
within those parts.
The survey light curves $\pounds_g(L)$ are computed for (parts of) the 297
survey galaxies in different observations with the same procedures.
The total survey light curve $\pounds(L)$ for a group of galaxies is
computed by summing up the survey light curves of individual galaxies in their
respective observations.
%

\subsection{Statistical descriptions of ULX samples}

With the contamination and the survey blue light computed for a survey,  we
study the statistical properties of the ULX population extracted from this
survey by calculating the fractions of survey galaxies that host ULXs, the ULX
rates per galaxy, the ULX rates per unit stellar light, and the luminosity
function.
The calculations are carried out for ULXs above different luminosities to study
the evolution with luminosity.


Some basic quantities are calculated for a survey of galaxies.
For a luminosity bin $(L_1,L_2)$, we calculate the observed number of ULXs
$U_g(L_1,L_2)$ in each galaxy and the observed total number of ULXs
$U(L_1,L_2)$ in all galaxies, the predicted number of contaminating sources
$N_b^g(L_1,L_2)$  for each galaxy and the total number $N_b(L_1,L_2)$ for all
galaxies, the survey blue light curve $\pounds_g(L)$ for each galaxy and the
total survey blue light curve $\pounds(L)$ for all galaxies.
The net number of ULXs for a survey in a luminosity bin $(L_1,L_2)$ is
calculated as $N_t(L_1,L_2) = U(L_1,L_2) - N_b(L_1,L_2)$, with the nominal
$1\sigma$ error as the maximum of $\sqrt{U(L_1,L_2)}$ and unity.
The corresponding cumulative quantities, the observed cumulative number of ULXs
$U($$>$$L_1)$, the cumulative number of contaminating sources $N_b($$>$$L_1)$,
the cumulative net number of ULXs $N_t($$>$$L_1)$ and its error, are calculated
with $L_2$ set to $10^{42}$ erg/sec.


The occurrence frequency, i.e., the fraction of survey galaxies that host ULXs
with luminosities above $L$ is calculated by comparing the number of survey
galaxies for which ULXs with luminosities above $L$ can be detected and the
number that host ULXs with luminosities above $L$.
The former, $N_{Sg}($$>$$L)$, is defined as the number of galaxies for which at
least 10\% of the blue light $\pounds_g$ have detection thresholds below $L$.
The latter, $N_{Hg}($$>$$L)$, is defined as the number of galaxies for which
$U_g(>\!L) - N_b^g(>\!L) > 0.95$, i.e., a galaxy with one ULX candidate and
$<$0.05 contaminating sources is considered as a host galaxy.
With these two numbers, the occurrence frequency is calculated as the minimum
of 100\% and $N_{Hg}($$>$$L)$/$N_{Sg}($$>$$L)$.

The ULX rates per survey/host galaxy are calculated for ULXs with luminosities
above $L$.
The ULX rate per survey galaxy is calculated as
$N_t^S($$>$$L)$/$N_{Sg}($$>$$L)$ with $N_t^S($$>$$L)$ as the net number of ULXs
in these $N_{Sg}($$>$$L)$ survey galaxies. The error for $N_t^S($$>$$L)$ is the
maximum of unity and $\sqrt{U^S(\!>L)}$ with $U^S(>L)$ as the observed number
of ULXs in these survey galaxies.
Similarly, the ULX rate per host galaxy is calculated as
$N_t^H($$>$$L)$/$N_{Hg}($$>$$L)$ with $N_t^H($$>$$L)$ as the net number of ULXs
in the $N_{Hg}($$>$$L)$ host galaxies. The error for $N_t^H($$>$$L)$ is the
maximum of unity and $\sqrt{U^H(\!>L)}$ with $U^H(>L)$ as the observed number
of ULXs in these host galaxies.
%


The ULX rate is also calculated as the number of ULXs per unit blue stellar
light to account for the variable sizes of survey galaxies.
For a bin $(L_1,L_2)$, the rate is calculated as $$R(L_1,L_2) = \sum_{L_1 \le
L_i \le L_2} {1 \over \pounds(L_i)} - \int_{L_1}^{L_2} {dN_b(L,L+dL) \over
\pounds(L)}$$
Here the sum is over all observed ULXs in the bin, and gives the observed ULX
rate; the integral gives the contamination rate, with $dN_b(L,L+dL)$ as the
number of contaminating sources in the luminosity interval $(L,L+dL)$.
The error $E(L_1,L_2)$ is computed as $$E^2(L_1,L_2) = \sum_{L_1 \le L_i \le
L_2} {1 \over \pounds^2(L_i)}$$
If there are no observed ULXs in the bin, the error is defined as $E(L_1,L_2) =
1/\bar\pounds(L_1,L_2)$ with $\bar\pounds(L_1,L_2) = \int_{L_1}^{L_2}
\pounds(L) dlgL/\int_{L_1}^{L_2} dlgL$ as the average of $\pounds(L)$ in the
$lgL$ space.
If the survey light curve is a  constant of $\pounds$ in the bin, the above
degenerates to $R(L_1,L_2) = {U(L_1,L_2)-N_b(L_1,L_2) \over \pounds}$ and
$E(L_1,L_2) = {\sqrt{U(L_1,L_2)} \over \pounds}$.
For a bin $(L_1,L_2)$, the cumulative ULX rate $R($$>$$L_1)$ and its error
$E($$>$$L_1)$ are computed using the above expressions with $L_2$ set to
$10^{42}$ erg/sec.

The above quantities are calculated for $L=10^{38}$ erg/sec to $L=10^{41}$
erg/sec in steps of $\Delta lgL = 0.2$.
The quantities for some luminosities are tabulated to illustrate how they
evolve with luminosity.
In Table 1 the occurrence frequencies and ULX rates
per galaxy are listed for ULXs above $lgL>39/39.2/39.6/40$ in different surveys
in our study.
For each luminosity, we list (1) the number of survey galaxies, (2) the number
of host galaxies, (3) the fraction, (4) the net number of ULXs per host galaxy,
and (5) the net number of ULXs per survey galaxy.
The ULX rates per unit stellar light for different surveys are listed in Table
2 for bins $lgL=[39,39.2]$ and $lgL=[39.2,39.4]$, and in Table 3 for bins
$lgL=[39.6,39.8]$ and $lgL=[40,40.2]$.
For each bin, we list (1) the observed number and the observed cumulative
number of ULXs, (2) the number and the cumulative number of contaminating
sources, (3) the net number of ULXs and its error, (4) the cumulative net
number of ULXs and its error, (5) the average blue light in the bin,  (6) the
ULX rate and its error, and (7) the cumulative ULX rate and its error.


The luminosity function of X-ray point sources is calculated as the ULX rate
per unit stellar light as a function of luminosity, in the luminosity range
from $10^{38}$ erg/sec to $10^{41}$ erg/sec.
Such a luminosity range covers ULXs and the luminous end for high-mass X-ray
binaries (HMXBs) and low-mass X-ray binaries (LMXBs) commonly seen in our
Galaxy and nearby galaxies, and enables direct comparisons between ULXs and
HMXB/LMXBs.
The luminosity function is fitted to the form of $dn/dL = \alpha L^{-\beta}$,
i.e., $dn/dlgL = \alpha L^{-\beta+1}$,
by minimizing $$\chi^2(\alpha,\beta) \equiv \sum_i {(N_t(L_i,L_{i+1}) -
N_f(L_i,L_{i+1}))^2 \over \sigma_i^2}$$
where $N_t(L_i,L_{i+1})$ is the net number of sources in the bin
$(L_i,L_{i+1})$,
$N_f(L_i,L_{i+1}) = \int_{L_i}^{L_{i+1}} \alpha L^{-\beta} \pounds(L) dL $ is
the number of sources predicted by the power-law model, 
and $\sigma_i$ is the error as the maximum of unity and
$\sqrt{U(L_i,L_{i+1})}$.
The best fits are computed for ULX populations in some groups of galaxies.  For
each fit, the luminosity interval for the fit, $\alpha$, $\beta$ with their
68.3\% errors, and the reduced $\chi^2$ are listed in Table 4.

\section{PROPERTIES OF DIFFERENT ULX SAMPLES}

With the methods described in Section 2, here we discuss the statistical
properties for the  ULX sample from all survey galaxies, and samples from
different groups of galaxies to reveal possible connections between the ULX
phenomenon and the galaxy properties.

\subsection{ULXs in all galaxies}


The total HRI survey includes 313 galaxies, about 7.1\% of the RC3 galaxies
within 40 Mpc.
Compared to the RC3 galaxies, the survey galaxies are slightly larger, closer,
optically or X-ray brighter.
In spite of slight over-abundances of ellipticals, lenticulars and S0/a--Sc
early spirals, and an under-abundance of dwarf spirals and irregulars, the
survey galaxy sample is representative of the nearby galaxies by morphological
types.
The ULX sample extracted from these galaxies is thus expected to be
representative of surveys without large biases in galaxy types, and its
properties can be used as standards to compare to properties of ULX samples
from different groups of galaxies.
%


A deep survey is constructed from observations of all galaxies with the
contamination from background/foreground objects carefully calculated following
the procedures in Section 2.
This surveys a total sky of 1.65 squared degrees and a total blue light of
$2.3\times10^{12} L_\odot$, with an average sensitivity of about
$2\times10^{-14}$ erg/sec/cm$^2$ in flux and $4\times10^{39}$ erg/sec in
luminosity.
With the $logN$--$logS$ relation, 26 contaminating sources with apparent
luminosities above $10^{39}$ erg/sec are predicted for the surveyed area within
the isophotes of galaxies, about 29\% of the 89 detected ULXs.
The contamination rate is 30\% (20.3/67) for ULXs above $1.6\times10^{39}$
erg/sec, 26\% (9/26) for ULXs above $4.0\times 10^{39}$ erg/sec, and 29\%
(2.6/9) for ULXs above $10^{40}$ erg/sec.


The occurrence frequencies and ULX rates are calculated with the procedures in
section 2.4 and tabulated in Table 1--3.
In the deep survey, there are 98 galaxies that could have a ULX above $10^{39}$
erg/sec detectable, and 35 galaxies (i.e., 36\% of the 98 galaxies) host a net
number of $>0.95$ ULXs.
This fraction is 25\%  (33/131) for ULXs above $1.6\times 10^{39}$ erg/sec,
8.8\% (20/226) for ULXs above $4.0\times 10^{39}$ erg/sec, and 3\% (8/281) for
ULXs above $10^{40}$ erg/sec.
The number of ULXs per survey galaxy is
$0.56\pm0.08$/$0.32\pm0.06$/$0.10\pm0.02$/$0.02\pm0.01$ for ULXs above
1/1.6/4/10$\times 10^{39}$ erg/sec.
The ULX rate per $10^{10} L_\odot$ is
$0.59\pm0.10$/$0.33\pm0.06$/$0.13\pm0.03$/$0.03\pm0.01$ for ULXs above
1/1.6/4/10$\times10^{39}$ erg/sec.
Given the average of $\sim0.8\times10^{10} L_\odot$ per survey galaxy, the ULX
rates per $10^{10} M_\odot$ are consistent with the ULX rates per survey
galaxy.

The occurrence frequencies can be compared to those given by Ptak \& Colbert
(2004).
In their HRI survey for ULXs in nearby galaxies, they compute the luminosities
in 2--10 keV using a power-law spectrum with a photon index of 1.7, and the
resultant luminosities are about 1.5 times lower than the luminosities computed
in 0.3--8 keV in this survey.
Ptak \& Colbert (2004) find that the fraction of the survey galaxies that host
more than one ULX is 12.3\% for ULXs above  $10^{39}$ erg/sec, 7.3\% for ULXs
above $2\times10^{39}$ erg/sec, 3.1\% for ULXs above $5\times10^{39}$ erg/sec,
and 1.1\% for ULXs above $10^{40}$ erg/sec.
These fractions are up to two times lower than the fractions computed from this
survey after proper considerations of the difference in the luminosities.
Note, however, that their criteria on whether a galaxy host a ULX are different
from ours and tend to underestimate the number of host galaxies.
For example, a galaxy is said to host ULXs in this survey if there are one
detected ULX and 0.04 predicted contaminating sources, while it is not by
their criteria.


An average survey is also computed for all galaxies in comparison to the deep
survey.
In the average survey, the occurrence frequency is 40\% (34/84) for ULXs above
$10^{39}$ erg/sec, 28\% (32/113) for  ULXs above $1.6\times 10^{39}$ erg/sec,
8.9\% (19/212) for ULXs above  $4\times 10^{39}$ erg/sec, and 3.8\% (10/271)
for ULXs above  $10^{40}$ erg/sec.  
The ULX rate is $0.52\pm0.09$/$0.32\pm0.08$/$0.09\pm0.02$/$0.03\pm0.01$ ULXs
per survey galaxy, and $0.70\pm0.12$/$0.38\pm0.07$/$0.11\pm0.03$/$0.04\pm0.02$
ULXs per $10^{10} L_\odot$ for ULXs above 1/1.6/4/10$\times 10^{39}$ erg/sec.
These occurrence frequencies and ULX rates are consistent with those in the
deep survey; we will not list them for the surveys of galaxies in the following
subsections.


The luminosity function of the X-ray point source populations, scaled by the
surveyed blue light, is calculated in the luminosity range of
$10^{38}$--$10^{41}$ erg/sec (Figure 1).
It can be well fitted with a power-law form $dn/dL = 0.56^{+0.19}_{-0.19}
L^{-1.96^{+0.40}_{-0.29}}$ per $10^{10} L_\odot$ in the luminosity range of
2.5$\times 10^{38}$ -- $10^{40}$ erg/sec (Table 4).  
At luminosities below $2\times10^{38}$ erg/sec, the luminosity function drops
from the power-law fit due to incompleteness.
Despite the artificial  classification of X-ray sources into HMXB/LMXBs and
ULXs, there are no breaks, i.e., no changes of the slope or the normalization
in the luminosity function around $10^{39}$ erg/sec between HMXB/LMXBs and
ULXs.
If the ULXs and HMXB/LMXBs are drawn from different binary distributions with
different primary masses, binary properties and accretion rates, breaks in the
combined luminosity function should be expected.  The absence of such breaks,
therefore, implies ULXs below $10^{40}$ erg/sec and the HMXB/LMXBs are drawn
from the same binary distribution in spite of their different luminosity
ranges.

%

A break in the luminosity function is present around $10^{40}$ erg/sec with
marginal significance, which separates the regular ULXs below $10^{40}$ erg/sec
and the extreme ULXs above $10^{40}$ erg/sec.
For the bin at $10^{40}$ erg/sec, only one ULX is detected while the power-law
fit would predict 5.0 true ULXs, and the $logN$--$logS$ relation would predict
1.4 contaminating sources. 
The deficit of ULXs in this bin is on a $2.4\sigma$ significance level, if the
number of ULXs in this bin has a $1\sigma$ fluctuation of $\sqrt{5}$.
As compared to the luminosity function for HMXB/LMXBs and the regular ULXs that
decreases with the luminosity, the luminosity function flattens after the gap,
though the ULX rates for the extreme ULX population are consistent with the
power-low fit within the $1\sigma$ errors except for in the gap.
For luminosities above $10^{40}$ erg/sec, nine extreme ULXs are detected, 2.6
contaminating sources are predicted, and 15.0 sources are expected from the
power-law fit.
The deficit of extreme ULXs is on a $2.2\sigma$ significance level, if the
number of extreme ULXs has a $1\sigma$ fluctuation of $\sqrt{15}$.
Such a break around $10^{40}$ erg/sec is expected as to separate the stellar
mass X-ray binaries and intermediate mass X-ray binaries (Grimm et al. 2003).
However, the significance for this break is marginal, and larger surveys with
more ULXs would be required to test whether it is a true feature.
%

\subsection{ULXs in galaxies of different types}

In our analysis, the galaxies are divided into early-type galaxies, late-type
galaxies and peculiar galaxies, and further into sub-types as elliptical,
lenticular, Sa, Sb, Sc, Sd, and Sm galaxies based on the Hubble type T (Table
5).
The survey galaxies cover a complete range of morphological types, and provide
a chance to study the ULX populations in different types of galaxies with a
uniform data set.


The contamination fraction of the ULX sample from the early-type galaxies is
much higher than that from the late-type galaxies (Tables 2--3).
In the deep survey of the early-type galaxies, 10 ULXs above $10^{39}$ erg/sec
are detected, while 10.3 contaminating sources are predicted for its surveyed
area of 0.33 squared degrees. 
For the late-type galaxies, 79 ULXs above $10^{39}$ erg/sec are detected, while
15.7 contaminating sources are predicted for its surveyed area of 1.32 squared
degrees.
The contamination fractions indicate that 63 (80\%) of the ULXs from the
late-type galaxies are true ULXs, and the ULX sample from the early-type
galaxies is dominated by contaminating sources.
%


A significant preference for ULXs to occur in the late-type galaxies is clearly
shown in the statistical results of the occurrence frequencies and ULX rates
(Tables 1--3).
%
%
The occurrence frequency for ULXs above $10^{39}$ erg/sec is 45\% (34/76) in
the late-type galaxies, 10 times higher than the 4.5\% (1/22) in the early-type
galaxies.
For ULXs above 1.6/4/10$\times 10^{39}$ erg/sec, the occurrence frequencies are
31.2\%/10.7\%/3.1\% in the late-type galaxies, much higher than the fractions
of 8.6\%/4.5\%/2.3\% in the early-type galaxies, though to lesser degrees for
ULXs with higher luminosities.
%
%
The ULX rates in the late-type galaxies are significantly higher than those in
the early-type galaxies for the regular ULXs.
For ULXs above $10^{39}$ erg/sec, for example, the ULX rate per survey galaxy
in the late-type galaxies ($0.72\pm0.11$) is at least 5 times higher than that
in the early-type galaxies ($0.02\pm0.11$).
For the extreme ULXs above $10^{40}$ erg/sec, the ULX rates in the late-type
galaxies are still higher than those in the early-type galaxies, but they are
consistent with each other within errors.
Note that the ULX rates in the early-type galaxies are consistent with being
zero, which reflects the equity between the number of detected ULXs and the
number of contaminating sources.


The X-ray sources in the early-type galaxies are dominated by the LMXB
population as seen from the luminosity function (Figure 2).
Due to the lack of young massive stars in the early-type galaxies, most of the 
X-ray binary sources are LMXBs.
In the deep (average) survey, the LMXB population shows a sharp cut-off at
$10^{39}$ ($1.6\times10^{39}$) erg/sec, which presumably corresponds to the
Eddington luminosity of a $\sim10M_\odot$ black hole.
The rates of LMXBs in the early-type galaxies are much higher than those in the
late-type galaxies and those from the power-law fit to the total X-ray
population in all galaxies.
In comparison, the ULX rates are low and consistent with being zero within
$1\sigma$ errors for most luminosity bins; this reflects the equity between the
number of detected ULXs and the number of contaminating sources.
No power-law is fitted to the luminosity function of the ULX population
separately due to the small number of detected ULXs in the early-type galaxies.
%


For the late-type galaxies, the luminosity function can be fitted with a
power-law form of $dn/dL = 0.62^{+0.21}_{-0.21} L^{-1.73^{+0.36}_{-0.25}}$ per
$10^{10} L_\odot$ in the luminosity range of 2.5$\times 10^{38}$ -- $10^{40}$
erg/sec (Figure 3).
This fit has a shallower slope than the power-law fit to the total X-ray
population in all early-type and late-type galaxies, because the late-type
galaxies have a significant ULX population while there is no significant ULX
population in the early-type galaxies.
Unlike in the early-type galaxies, there are no breaks in the luminosity function
between the HMXB/LMXBs and the regular ULXs, indicating that the regular ULXs
and the HMXB/LMXBs may be drawn from the same binary population in the
late-type galaxies.
Similar to the luminosity function from all galaxies, there is marginal
evidence that the regular ULX population is different from the extreme ULX
population, because of the flattening of the luminosity function for extreme
ULXs, the deficit of extreme ULXs (significant on a $3.2\sigma$ level), and the
gap (significant on a $2.4\sigma$ level) separating the two ULX populations.


The statistical properties of ULX samples from different sub-types of galaxies
are calculated (Tables 1--3) to reveal possible trends with galaxy sub-types.
The occurrence frequencies and ULX rates generally increase with the Hubble
type T until they peak at Sc, then decrease with T (Figures 4--5).
For Sc galaxies, the occurrence frequency (the ULX rate per survey galaxy) is
59\% ($1.10\pm0.21$) for  ULXs above $10^{39}$ erg/sec, much higher than those
for the average late-type galaxies and the early-type galaxies.
When scaled by the survey blue light, the ULX rates increases monotonically
with T, and the rates in Sm spirals are highest among the sub-types of galaxies
due to the small sizes of Sm spirals ($0.077 \times 10^{10} L_\odot$ per Sm
galaxy as compared to $0.94\times 10^{10} L_\odot$ per Sc galaxy).


The results of this HRI survey can be compared to the results of the recent
Chandra ACIS survey of ULXs in 82 nearby galaxies (Swartz et al. 2004).
For the Chandra survey, there are $\sim$2 ULX candidates per galaxy for both
early-type galaxies and late-type galaxies. The number of ULXs per unit
($10^{42}$ erg/sec) blue luminosity is $0.11\pm0.02$ ($0.30\pm0.11$) for
early-type (late-type) galaxies, or $0.55\pm0.10$ ($1.50\pm0.55$) ULXs per
$10^{10} L_\odot$ given $L_{B\odot} \sim 5\times 10^{32}$ erg/sec.
These rates are higher than the rates from the HRI survey, especially for
early-type galaxies. 
One reason for the rate difference lies in the different sensitivity ranges for
HRI observations (0.5--2.4 keV) and ACIS observations (0.3--10 keV), which
enables ACIS observations to detect highly absorbed ULXs buried in dust spiral
arms.  This is important given that $\ge75$\% of the ULXs (Paper I) are
associated with spiral arms or other dust features.
Another possible reason is the source blending effect, which is easier to occur
in HRI observations for the poorer spatial resolution ($\sim5^{\prime\prime}$) than
in ACIS observations ($\sim1^{\prime\prime}$).
When the surface source density exceeds what the telescope can resolve, sources
blend to reduce the number of sources and increase the diffuse X-ray
background. 
A higher diffuse background increases the detection threshold, and lowers the
luminosities of detected sources, both of which further reduce the number of
detected sources.

\subsection{ULXs in starburst and non-starburst galaxies}

Many ULXs have been found to reside in star forming regions, which leads to the
postulation that ULXs are closely related to star formation activities.
The statistically significant preference for ULXs to occur in late-type
galaxies as reported in the previous subsection is consistent with this
postulation, since late-type galaxies are known to have star formation
activities while few early-type galaxies have such activities.
To study the connection between star formation and the ULX phenomenon, we group
galaxies based on whether they are starburst galaxies that have significant
{\it current} star formation activities.
While no single definition of the starburst phenomenon exists, the survey
galaxies listed as starburst or HII galaxies in NED are included in the
starburst galaxy group (Sbrst), which include 46 out of 313 survey galaxies,
with two E0 ellipticals, three dwarf lenticulars and 41 late-type galaxies.
For comparison also constructed is another group of 164 non-starburst late-type
galaxies (nSbrstL) that have milder star formation activities.
We caution that the list of starburst/HII galaxies is not complete, and some
galaxies in the non-starburst galaxy group may be starburst galaxies.

The contamination fraction of the ULX sample from the starburst galaxies is
much lower than the fractions from non-starburst galaxies.
In the deep survey of the starburst galaxies, 3.1 contaminating sources are
predicted, while 35 ULXs above $10^{39}$ erg/sec are detected.
The contamination fraction (9\%) is much lower than that for non-starburst
late-type galaxies (29\%), and that for early-type galaxies (100\%).
The lowest contamination fraction implies that the starburst galaxies have the
highest ULX surface density as compared to the other groups of galaxies, given
that the number of contaminating sources is determined by the surface areas of
the galaxies.


The occurrence frequencies and ULX rates in the starburst galaxies are higher
than those in other types of galaxies.
For ULXs above $10^{39}$ erg/sec, the occurrence frequency in starburst
galaxies is 69\% (18/26), much higher than that in the non-starburst late-type
galaxies (31\%) and that in the early-type galaxies (4.5\%).
The ULX rates per survey galaxy (per $10^{10} L_\odot$) in the starburst
galaxies are $0.98\pm0.20$ ($1.5\pm0.29$), about two times higher than those in
the non-starburst late-type galaxies ($0.56\pm0.12$/$0.57\pm0.13$).
For the extreme ULXs above $10^{40}$ erg/sec, the occurrence frequencies and
ULX rates in the starburst galaxies are still higher than those in other types
of galaxies, though to a lesser degree, and they are consistent with being the
same within $1\sigma$ errors.


The difference of the ULX rates in starburst galaxies and non-starburst
late-type galaxies can be used to estimate the ULX fractions due to {\it
current} star formation.
For ULXs above $10^{39}$ erg/sec, the difference is $0.93\pm0.42$ ULXs per
$10^{10} L_\odot$, about 1.5 times of the ULX rate in the non-starburst
late-type galaxies.
This difference implies that about 60\% of the ULXs in the starburst galaxies
are linked to current star formation, if the difference is attributable
exclusively to the current star formation. 
Note that this fraction is a lower limit since some of the ``non-starburst''
galaxies have high star formation rates as noted in the Section 3.4, and
and  some of the ULXs may be linked to the star formation activities in these
galaxies.
For the extreme ULXs above $10^{40}$ erg/sec, the ULX rates in the starburst
galaxies are higher than the rates in the non-starburst galaxies, however, only
by an amount that is consistent with being zero within errors.


The luminosity function for the X-ray source population in the starburst
galaxies (Figure 6) can be fitted with a power-law form of $dn/dL =
1.08^{+0.47}_{-0.47} L^{-1.73^{+0.68}_{-0.35}}$ per $10^{10} L_\odot$ in the
luminosity range of of $4\times 10^{38}$ -- $10^{40}$ erg/sec. 
There is no significant break between the HMXB/LMXB population and the regular
ULX population, except for an insignificant deficit of sources in the
luminosity bin of 6.3--10 $\times10^{38}$ erg/sec.
The luminosity bin of 1.6--2.5$\times10^{39}$ erg/sec is significantly lower
than the best fit in the deep survey, yet may be a fluctuation in the counts,
since the bin is consistent with the fit in the average survey.
Similar to the luminosity function from the late-type galaxies, there is
marginal evidence that the regular ULX population is different from the extreme
ULX population, because of the flattening of the luminosity function for
extreme ULXs, the deficit of extreme ULXs (significant on a $1.5\sigma$ level),
and the gap (significant on a $1.6\sigma$ level) separating the two ULX
populations.
This luminosity function is consistent with the universal luminosity function
derived from Chandra observations of nearby starburst galaxies (Grimm et al.
2003), in that both show a similar power-law slope ($\sim1.6$), a cutoff around
$10^{40}$ erg/sec, and no breaks around $10^{39}$ erg/sec.

The above power-law fit has a shallower slope and a normalization factor 2
times higher as compared to the power-law fit for the X-ray source population
in the non-starburst late-type galaxies (Figure 7).
The higher normalization indicates that the starburst galaxies have higher ULX
rates than the non-starburst galaxies.
The shallower slope indicates that the starburst galaxies have relatively more
high luminosity ULXs than the non-starburst galaxies. 
However, this implication is not conclusive since the two slopes are consistent
within the large fitting errors.

\subsection{ULXs in galaxies with different star formation rates}


As discovered in comparing ULXs in starburst and non-starburst galaxies, there
are about two times more ULXs per $10^{10} L_\odot$ in starburst galaxies than
in non-starburst galaxies.
To quantify the relation between star formation and the ULX phenomenon, we
group the galaxies based on their star formation rates.
The star formation rate is calculated with ${\rm SFR}(M_\odot/yr) =
4.5\times10^{-44}\times L(60\mu)$ (Rosa-Gonzalez et al. 2001). The flux
densities at $60 \mu$, listed in Table 1 of Paper I (Liu \& Bregman 2005), are
taken from the IRAS point source catalog (IPAC, 1986), with some nearby
galaxies from Rice et al. (1988). For 102 galaxies that are not detected, the
$3\sigma$ upper limit is calculated by adopting noise levels of 8.5
mJy/arcmin$^2$ (Rice et al. 1988). 
The calculated rates are compared to the compilation of Grimm et al. (2003) for
11 galaxies, and are consistent with their rates within 50\% without systematic
biases.
Based on the calculated star formation rates, the galaxies are put into nine
groups, each with similar star formation rates (Table 5).


The occurrence frequencies and ULX rates increase with the star formation rates
of the galaxies (Tables 1--3). 
The occurrence frequency increases monotonically from $\sim10$\% for the group
SFRU ($<0.009 M_\odot/yr$) to 80\% for the group SFRD ($\sim8.3M_\odot/yr$) for
ULXs above $10^{39}$ erg/sec (Figure 8).
The general increase trend of ULX rates shows large scatters for low luminosity
ULXs, with a sharp drop from 1.5 ULXs above $10^{39}$ erg/sec per survey galaxy
for the group SFRC ($\sim2.1M_\odot/yr$) to 0.6 ULXs per survey galaxy for the
group SFRD(Figure 9).
This sharp drop is mainly attributable to the source blending effects, which
would occur and lead to fewer detected sources when the surface number density
exceeds what the X-ray telescope can resolve.
Comparison with Chandra observations shows that the blending effects reduce the
detected ULXs significantly in HRI observations for the galaxies in the group
SFRD.
For the 15 galaxies with both HRI and ACIS observations, there are only 7 ULXs
detected within the $D_{25}$ isophotes in the HRI survey, while there are
$\sim70$ ULXs detected in ACIS observations.
This number, 70 ULXs, exceeds the total numbers in the groups SFRC/H with
similar total star formation rates, implying that these two groups are also
affected by the blending effects, though to less degrees.


The general increase of ULX rates with the star formation rates tempts the
postulation of a linear relation between the two quantities. 
Grimm et al. (2003) tested the postulation with Chandra observations of 11
nearby starburst galaxies with ${\rm SFR}\ge 1M_\odot/yr$ in which the HMXBs
dominate the X-ray sources, and found a universal luminosity function, i.e.,
${dN \over dL_{38}} = 3.3^{+1.1}_{-0.8} {\rm SFR} L_{38}^{-1.61\pm0.12}$ for
$L_c<2\times10^{40}$ erg/sec, where $L_{38}$ is the X-ray luminosity in units
of $10^{38}$ erg/sec, and ${\rm SFR}$ is in units of $M_\odot/yr$.
There are no breaks in the universal luminosity function between HMXBs and ULXs
below $L_c$, indicating that the regular ULXs below $L_c$ are an extension of
the HMXB population in these starburst galaxies.
The number of such HMXB-like ULXs expected from a galaxy is thus linearly
proportional to the star formation rate of the host galaxy, i.e., $N_{ULX} =
\beta\cdot {\rm SFR}$.
Here $\beta\sim$1.8/0.4/0.1 for ULXs above 1/4/10$\times 10^{39}$ erg/sec.
Note that the luminosity is calculated in 2--10 keV in Grimm et al. (2003), and
is about 1.5 times lower than the luminosity in 0.3--8 KeV for a power-law
spectrum with $\Gamma = 1.7$.
%


The expected linear relation between the number of ULXs and the star formation
rate can be tested with the HRI survey.
The star formation rates of the survey galaxies span from $<0.001 M_\odot/yr$
to $15 M_\odot/yr$ with a median of 0.1 $M_\odot/yr$, and many ($>70$\%)
galaxies are expected to host less than one ULX.
To reduce the statistical fluctuation caused by the small number of ULXs in
individual galaxies, we sum up the number of ULXs with the predicted
contaminating sources subtracted and the star formation rate of each galaxy in
a group.
The linearity of the relation guarantees the same linear relation between the
total number of ULXs and the total star formation rate for a group of galaxies.
Here the total star formation rate (Tables 2--3) is calculated by summing up
the star formation rates of only galaxies that can have ULXs detected.
The expected linear relation is compared to the observed relation between the
total number of ULXs and the total star formation rate calculated for the HRI
survey (Figure 10).
{\it The observed relations for ULXs above 1/4/10$\times10^{39}$ erg/sec cannot
be fitted with the expected linear relations, indicating that  the ULXs are not
a mere extension of the HMXB population}.
%


A tight non-linear correlation is shown between the total number of ULXs above
$10^{39}$ erg/sec ($N(L>10^{39})$) and the total star formation rate. 
This can be fitted with  $N(L_X>10^{39}) = 7.0^{+1.9}_{-1.9} {\rm
SFR}^{0.36^{+0.07}_{-0.10}}$; the group SFRD is excluded from the fit for
severe blending effects.
For the galaxies with high star formation rates ($\ge1M_\odot/yr$, in groups
SFRC/D/G/H), there are fewer ULXs detected than expected HMXB-like ULXs as an
extension of the HMXB population, which may be attributed to the blending
effects in these galaxies.
For the galaxies with low star formation rates ($<0.4 M_\odot/yr$, in groups
SFRU/A/B/E/F), the expected HMXB-like ULXs are significantly fewer than the
detected ULXs. 
These extra ULXs are an extension of the LMXB population in these galaxies
given the dominance of LMXBs and the absence of breaks in the luminosity
function between the LMXB and ULX populations.
%
%
The HMXB-like ULXs correspond to the young X-ray point source population
associated with the young stellar population (Pop I) as in Colbert et al.
(2004), and the LMXB-like ULXs correspond to the old X-ray point source
population associated with the old stellar population (Pop II).


The relative fractions of LMXB-like ULXs and HMXB-like ULXs change with the
star formation rates of the galaxies.
If the number of LMXB-like ULXs is linearly proportional to the total light and
the the number of HMXB-like ULXs to the total star formation rate, the total
number of ULXs can be expressed as $N = \alpha \cdot \pounds + \beta \cdot
SFR$.
%
%
%
Given the surveyed light and star formation rates, and the detected ULXs with
the contaminating sources removed for groups SFRA and SFRC (Table 2), we get
$\alpha = 0.93$ and $\beta = 0.20$ for ULXs above $10^{39}$ erg/sec.
This relation underestimates the number of HMXB-like ULXs than
predicted from the linear relation of Grimm et al (2003) due to the difference
between HRI and ACIS in the sensitivity ranges, and the source blending effects
at high star formation rates.
For example, this relation predicts the group SFRD should have 15 ULXs, while
there are $\ge 70$ ULXs detected with Chandra observations.
Nevertheless, this relation predicts correctly the trend that the LMXB-like
ULXs dominate for low specific star formation rate ($\ll 4.6
M_\odot/yr/10^{10}L_\odot$), and the HMXB-like ULXs dominate for high specific
star formation rate ($\gg 4.6 M_\odot/yr/10^{10}L_\odot$).


The total number of ULXs above $4\times10^{39}$ erg/sec can be fitted with
$N(L_X>4\times10^{39}) = 3.5^{+1.7}_{-1.7} {\rm SFR}^{0.13^{+0.13}_{-0.19}}$.
This has a shallower slope than the fit for ULXs above $10^{39}$ erg/sec, and
implies a steeper slope for the luminosity function of ULXs in galaxies with
higher star formation rates.
There is a seeming anti-example for this expectation, i.e., the luminosity
function for ULXs in the starburst galaxies has a shallower slope than that in
the non-starburst galaxies.
This anti-example can be understood given the fact that, the starburst
galaxies, while characterized by significant current star formation activities,
have lower star formation rates than the non-starburst galaxies.
About 60\% (26/46) of the starburst/HII galaxies have rather low star formation
rates ranging from 0.002 to 1 $M_\odot/yr$ with a median of 0.06 $M_\odot/yr$.
In comparison, the non-starburst galaxies includes most of the galaxies with
highest star formation rates, including the two galaxies with the highest
rates, four of the six galaxies with $>10M_\odot/yr$, 14 of the 20 galaxies
with $>4M_\odot/yr$, and 27 of the 40 galaxies with $>2M_\odot/yr$.
%

\section{SPATIAL DISTRIBUTION OF ULXS}

To study the spatial distribution of ULXs and how ULXs trace the blue light, we
compare the ULX populations in elliptical annuli of the survey galaxies.
For each galaxy, the area within $2\times D_{25}$ is divided into twenty
elliptical annuli in steps of $0.1\times R_{25}$, with $R_{25}$ as the
elliptical radius of the $D_{25}$ isophote.
For a group of galaxies, we construct twenty ULX populations, with each
extracted from the annuli of the same elliptical radius range.
For each ULX population, the contamination and the survey blue light curve are
calculated excluding the $10^{\prime\prime}$ nuclear region as described in
Section 2.
Here we study the radial distributions for ULXs in late-type galaxies and
early-type galaxies.


There is a significant concentration of ULXs toward galactic centers in the
late-type galaxies.
The surface number density of ULXs above $10^{39}$ erg/sec, with the predicted
contaminating sources subtracted, decreases with radii until it flattens
outside the $D_{25}$ isophotes (Figure 11).
If the ULXs were all background and/or foreground objects, a random (i.e.,
flat) radial distribution would be expected.
The observed radial distribution, therefore, implies that more ULXs within the
$D_{25}$ isophotes are truly associated with the galaxies than the ULXs between
1--2 $\times D_{25}$.
In the deep survey of the late-type galaxies, there are 79 detected ULXs, 15.7
contaminating sources, and 63.3 ``true'' ULXs within the $D_{25}$ isophotes
(Table 2), while there are 60 detected ULXs, 43.1 contaminating sources, and
16.9 ``true'' ULXs between 1--2$\times D_{25}$.
Therefore the bulk of the ``true'' ULXs ($\sim80$\%) is within the $D_{25}$
isophotes.
Note that 12 ULXs between 1--2$\times D_{25}$ are associated with dust
loops/strips and spiral arms, and are apparently connected to the galaxies,
which account for $\sim70$\% of the ``true'' ULXs.


The radial distribution of ULXs is more extended than the blue light in the
late-type galaxies.
The survey blue light per annulus peaks at $R\sim h$ if the light follows an
exponential disk profile with a scale height of $h$ as described in Section
2.3. 
In the deep survey for the late-type galaxies, this peak occurs at $0.3R_{25}$
since $\bar{h}=0.3R_{25}$.
In comparison, the number of ULXs per annulus peaks at $0.5R_{25}$, indicative
of a higher scale height for the ULX distribution than that for the blue light
(Figure 12).
It is also suggested by the gradual increase in the number of ULXs per unit
blue light with radii (Figure 13).
This is expected since $\ge75$\% of the ULXs are on the spiral arms
(Paper I), while the spiral arms are more extended than the light profile.


For the early-type galaxies, the detected ULXs are radially distributed as the
predicted contaminating sources, except for a marginal excess of detected ULXs
around $0.4R_{25}$ (Figure 14) that vanishes for ULXs above $2.5\times10^{39}$
erg/sec.
Finding that the ULX candidates in early-type galaxies in CP2002 are
distributed randomly, Irwin et al. (2004) argue these candidates are
contaminating sources, and they contend that there are no true ULXs above
$2\times10^{39}$ erg/sec in early-type galaxies.
Indeed, there should be very few true ULXs in our survey of early-type
galaxies: 10 ULX candidates  are detected within the $D_{25}$ isophotes with
10.3 contaminating sources predicted, and 32 ULXs are detected between 1--2
$\times D_{25}$ with 29.7 contaminating sources predicted.

\section{DISCUSSIONS}


In this paper, statistical properties of the ULX populations constructed from
our recent HRI survey of X-ray sources in nearby galaxies (Liu \& Bregman,
2005) are studied with particular attention to the contamination problem.
The ULXs as extranuclear X-ray point sources above $10^{39}$ erg/sec are found
common in late-type galaxies. For example, $\sim$40\% of the late-type galaxies
harbor at least one ULX with $0.84\pm0.13$ ULXs per $10^{10} L_\odot$ in our
HRI survey.
The luminosity functions of the X-ray source populations in different galaxy
samples are analyzed to study the connection between HMXB/LMXBs and ULXs, and
to reveal the dependence of ULXs on the star formation activities.

\subsection{HMXB/LMXBs, regular ULXs and extreme ULXs}

This HRI survey has a significant coverage at low luminosities, with X-ray
sources of $2\times10^{38}$ erg/sec detectable in $\sim$10\% of the total
surveyed blue light.
The luminosity functions, therefore, are studied in the luminosity range of
$10^{38}$--$10^{41}$ erg/sec, which enables direct comparisons between the
ordinary HMXB/LMXB populations and the ULX populations.


The luminosity functions of the HMXB/LMXBs and ULXs in the late-type
galaxies can be fitted by a single power-law below $10^{40}$ erg/sec, with no
breaks between the two populations.
This is comparable to the work by Grimm et al. (2003), which shows that the
X-ray source (mainly HMXBs) populations in 12 nearby starburst galaxies, as
well as the Milky Way and Magellanic Clouds, follow the same universal
luminosity function scaled by the star formation rate, i.e., ${dN \over
dL_{38}} = 3.3^{+1.1}_{-0.8} {\rm SFR} L_{38}^{-1.61\pm0.12}$ for
$L_c<2\times10^{40}$ erg/sec, where $L_{38}$ is the X-ray luminosity in units of
$10^{38}$ erg/sec.
The absence of breaks between the ordinary HMXB/LMXBs and the regular ULXs
suggests that {\it the regular ULX population must be some kind of high
luminosity extension of the ordinary HMXB/LMXB populations}, all of which are
associated with the same underlying distributions of primary masses, binary properties,
and accretion rates.
The HMXB-like ULXs could be stellar mass black hole binaries with high mass
secondaries undergoing thermal timescale mass transfer through Roche lobe
overflow (King et al. 2001), and the LMXB-like binaries could be the soft X-ray
transients in their bright outburst phases (King et al. 2002).


The existence of a cutoff around $10^{40}$ erg/sec in the luminosity function
can have significant implications for understanding the nature of ULXs.
This cutoff, present both in Grimm et al. (2003) and in this work,  suggests a
cutoff of $\sim100 M_\odot$ in the mass distribution of black holes if the ULXs
radiate at the canonical Eddington luminosity. 
If the mass distribution of black holes cuts off around $\sim15M_\odot$, the
most massive black hole achievable in the ordinary stellar evolution models,
the ULXs must radiate at $>6$ times the Eddington luminosity.
These apparent super-Eddington radiation can result from many effects, such as
beamed emission (King et al. 2001), emission of a jet (Kording et al. 2002),
emission from a supercritical accretion disk (Shakura \& Sunyaev 1973), or
emission from a magnetized accretion disk where the photon-bubble instabilities
operate (Begelman 2002).


The extreme ULXs above $10^{40}$ erg/sec might be a different population from
the regular ULXs below $10^{40}$ erg/sec.
This is evidenced by the flattening of the luminosity function of the extreme
ULXs as compared to the luminosity function of the regular ULXs, and the
significant ($\sim2\sigma$) gap separating the two populations.
For the ULX population in the late-type galaxies, seven extreme ULXs are
detected, 1.5 contaminating sources are predicted, and 19.5 extreme ULXs are
expected from the power-law fit for the regular ULXs. These numbers suggest
that most of these extreme ULXs are ``true'' ULXs, and there are significantly
($\sim3\sigma$) fewer extreme ULXs than the regular ULXs.
The extreme ULXs are usually found in star-forming regions, and some have X-ray
spectral features of systems with black holes of $\ge1000M_\odot$, e.g., the
ULX in Holmberg IX (Miller et al. 2004) and the ULX in NGC1313 (Miller et al.
2003).
Thus it is reasonable to postulate that the extreme ULXs could be the
intermediate mass black holes as suggested by Colbert \& Mushotzky (1999),
while the regular ULXs are systems of stellar mass black holes.
There are, however, only a few extreme ULXs detected in the HRI survey, and
larger surveys with more extreme ULXs are needed to study whether the extreme
ULXs are a truly different population, and the details of the luminosity
function.


There is no significant ULX population in the early-type galaxies in our HRI
survey.
In the early-type galaxies, the X-ray sources are dominated by LMXBs, and the
luminosity function shows a cutoff at $\sim10^{39}$ erg/sec, suggestive of a
cutoff of $\sim10M_\odot$ in the mass distribution of stellar mass black holes
if they radiate at the Eddington luminosity.
Such a cutoff has been verified with Chandra observations of nearby elliptical
galaxies M87, M49 and NGC4697 (Jordan et al. 2004).
In the deep survey, 10 ULX candidates above $10^{39}$ are detected with 10.3
contaminating sources predicted from the $logN$--$logS$ relation, and the two
radial distributions are consistent with each other, indicative of absence of
``true'' ULXs in early-type galaxies.  This is consistent with the claimed lack
of ULXs above $2\times10^{39}$ erg/sec in the 0.3--10 keV band in early-type
galaxies (Irwin et al. 2004).
The ``true'' ULXs in early-type galaxies, if present, cannot be explained by
the thermal timescale mass transfer of HMXBs owing to the absence of high mass
stars; instead they can be soft X-ray transients in their bright outburst
phases (King et al. 2002).


\subsection{Star formation and the ULX phenomenon}


A close connection between the ULX phenomenon and star formation is revealed
statistically in the HRI survey.
This connection implies that ULXs preferentially occur in late-type galaxies
than in early-type galaxies owing to the lack of star formation in the latter;
it is indeed observed in the HRI survey.
In the deep survey, there are $0.72\pm0.10$ ULXs per survey galaxy for the
late-type galaxies, while only $0.02\pm0.10$ ULXs per survey galaxies for the
early-type galaxies.
We note that six of the ten ULX candidates in early-type galaxies are from the
lenticular galaxies NGC1316 and NGC5128 with recent merging events and induced
star formation activities.
In late-type galaxies, ULXs preferentially occur on the spiral arms or dust
lanes, with 75\% (81/106) of the ULXs in the clean sample of Paper I in such
regions.
This may explain why the radial distribution of ULXs has a larger scale height
than the light profile, since the spiral arms are more extended than the light.
The ULXs outside the $D_{25}$ isophotes consist of $\sim20$\% of all ULXs in
late-type galaxies, and are also preferentially found in star forming regions,
with about 70\%  associated with spiral arms or other dust regions.
Among late-type galaxies, the starburst/HII galaxies with significant current
star formation activities have higher ULX rates than the non-starburst
galaxies.
The close connection is further supported by a general increasing trend of the
occurrence frequencies and ULX rates with the star formation rate.


The close connection between ULXs and star formation has been confirmed by
recent optical observations with the Hubble Space Telescope and ground-based
telescopes.
Pakull \& Mirioni (2002) observed 15 ULXs in 11 galaxies, and found 13 ULXs
associated with well-defined HII regions and emission nebulae. 
In comparison, only a few ULXs are reported to associate with globular
clusters, e.g., the ULX in NGC4565 (Wu et al. 2001), and the two ULXs in
NGC1399 (Angelini et al.  2002).
Such a connection is consistent with the postulation that ULXs are high-mass
X-ray binary systems that are undergoing thermal timescale mass transfer through
Roche lobe overflow (King et al. 2002).
Unlike the ordinary Galactic HMXBs that accrete by capturing the stellar winds,
the accretion rates for these HMXB-like ULXs are high enough to power radiation
at near/super Eddington luminosities.
This HMXB scenario is directly supported by the optical identifications of some
ULXs with high mass stars, i.e., the ULX in M81 with an O8 V star (Liu et al.
2002), and the ULX in NGC5204 with a B0 Ib supergiant (Liu et al. 2004).


In addition to HMXB-like ULXs, there must be LMXB-like ULXs in the late-type
galaxies, since the HMXB-like ULXs alone cannot account for all the detected
ULXs.
The HMXBs trace the star formation activities for its short lifetime
($\sim10^7$ years), and are expected to be linearly proportional to the star
formation rate of a galaxy.
Grimm et al. (2003) verified this linear relation with Chandra observations of
nearby starburst galaxies, and found a universal luminosity function, i.e.,
${dN \over dL_{38}} = 3.3^{+1.1}_{-0.8} {\rm SFR} L_{38}^{-1.61\pm0.12}$ for
$L<2\times10^{40}$ erg/sec, where $L_{38}$ is the X-ray luminosity in units of
$10^{38}$ erg/sec.
If all ULXs are HMXB-like, a linear relation would be expected between the
number of ULXs and the star formation rates.
However, no such linear relation is found in this HRI survey, which includes
galaxies with star formation rates from $<0.001 M_\odot/yr$ to $17M_\odot/yr$.
Instead, a non-linear relation with the form $N(L_X>10^{39}) =
7.0^{+1.9}_{-1.9} {\rm SFR}^{0.36^{+0.07}_{-0.10}}$ is found for ULXs above
$10^{39}$ erg/sec.
For galaxies with low star formation rates ($<0.4M_\odot/yr$), the detected
ULXs are significantly more than the HMXB-like ULXs expected from the linear
relation.
This excess of ULXs can be contributed as LMXBs, which are soft X-ray
transients in their bright outburst phases as suggested by King et al. (2002).


The observed non-linear relation implies the fraction of HMXB-like ULXs
increases with the star formation rate, which is expected since the LMXB-like
ULXs trace the total mass/light instead of the star formation rate, thus its
rate keeps constant while the total ULX rate increases with the star formation
rate.
We estimate that the total number of ULXs $N = 0.93 \pounds + 0.20 SFR$, with
the surveyed light $\pounds$ in unit of $10^{10}L_\odot$, and the star
formation rate in unit of $M_\odot/yr$.
While this relation correctly predicts the increase of the HMXB-like ULXs with
the star formation rate, it cannot be taken quantitatively, since it severely
underestimates the contribution of HMXB-like ULXs owing to the difference
between HRI and ACIS in the X-ray spectral sensitivity and the source blending
effects at large star formation rates.
Chandra observations of galaxies with high star formation rates are needed to
alleviate the blending effects to accurately quantify the relative fractions of
HMXB-like ULXs and LMXB-like ULXs.


\acknowledgements

we are grateful to the NED, VizieR services. We would like to thank
Renato Dupke, Eric Miller, Rick Rothschild, and Steven Murray for helpful
discussions.  We gratefully acknowledge support for this work from NASA under
grants HST-GO-09073.

\clearpage



\begin{figure}
\plottwo{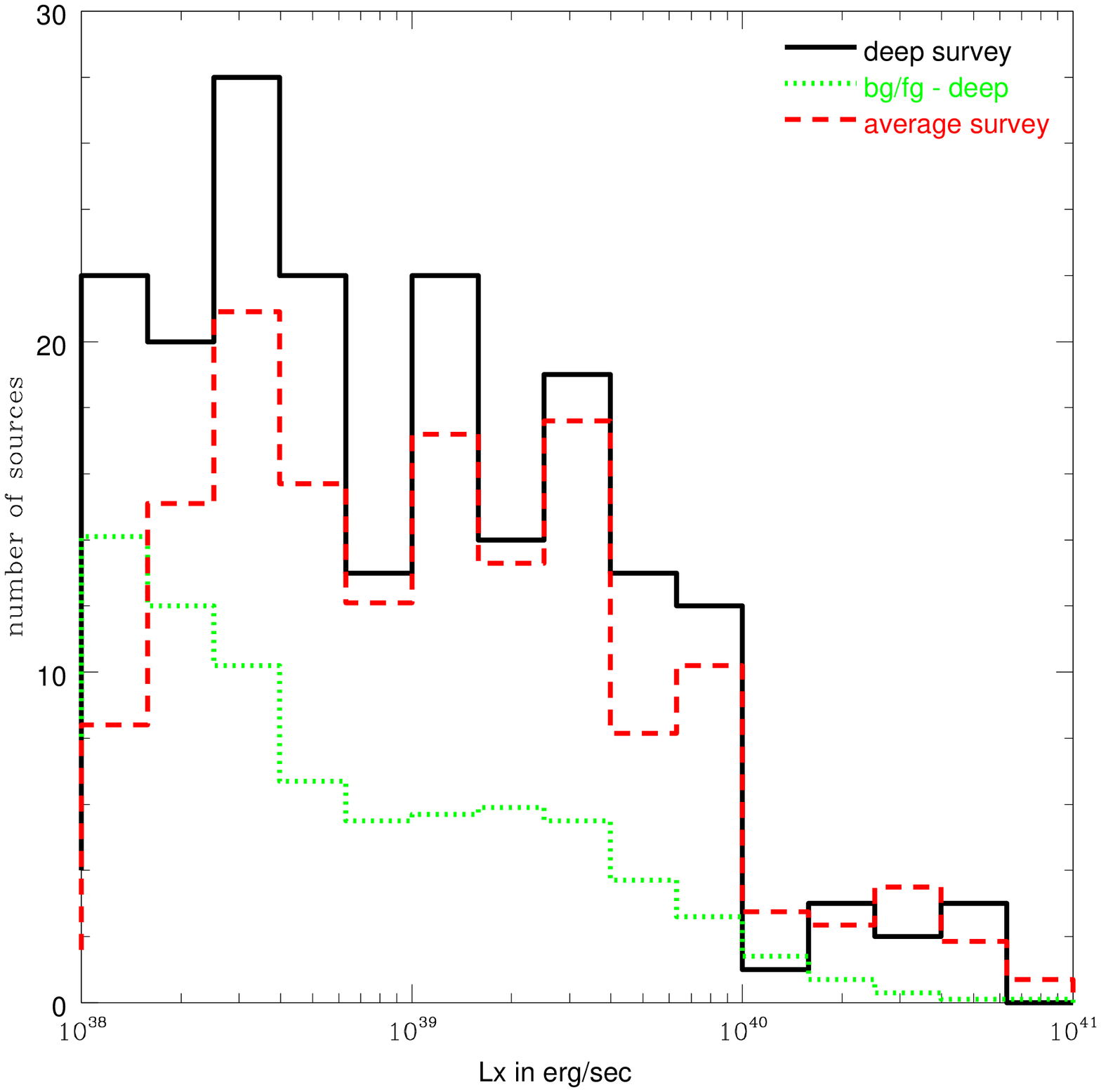}{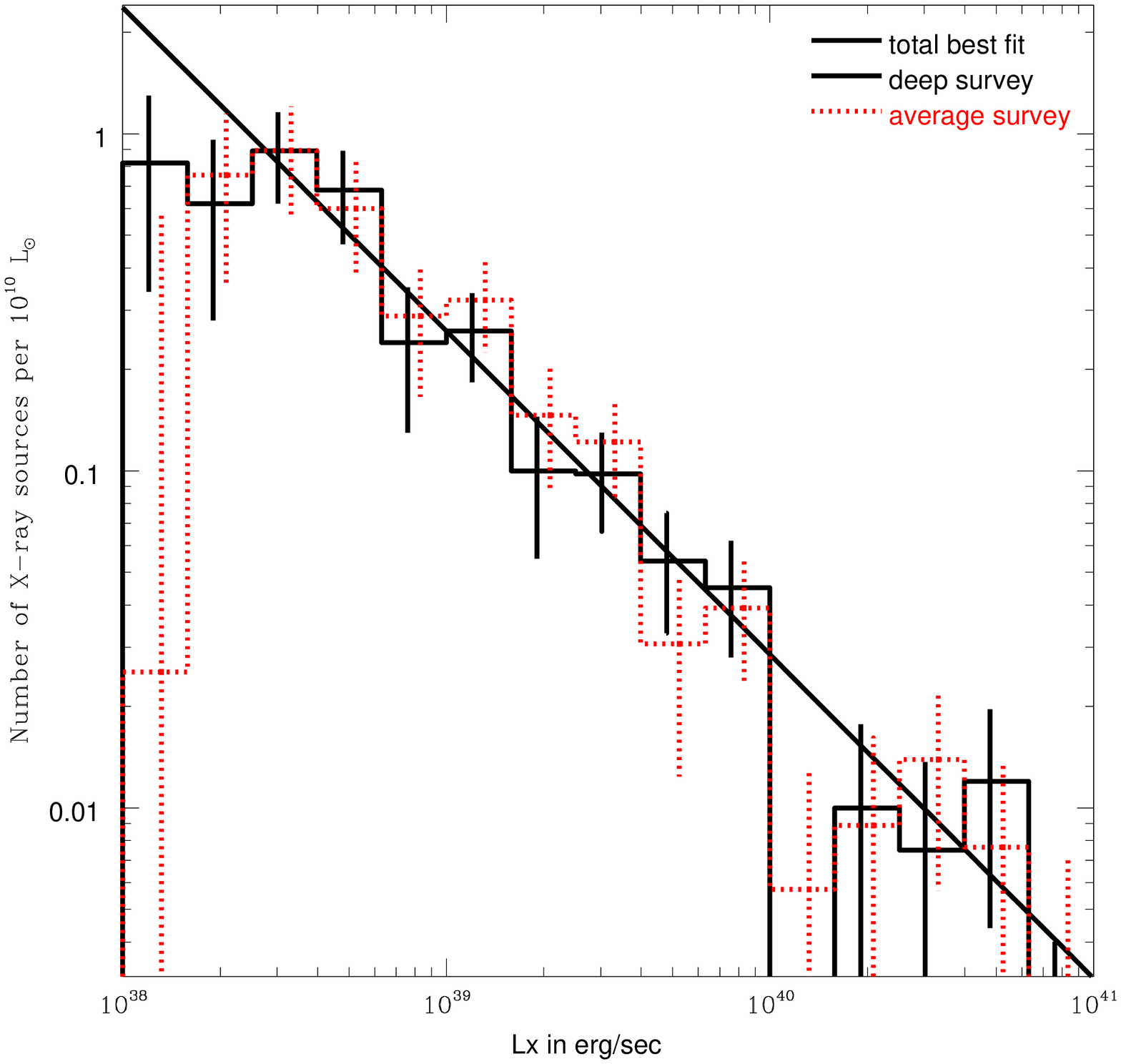}

\caption{The luminosity functions for the X-ray source population in all
galaxies in the deep survey and in the average survey. (left) Histograms of the
numbers of detected ULXs (solid) and contaminating sources (dotted) in the deep
survey, and detected ULXs in the average survey (dashed). (right) The
luminosity functions scaled by the survey blue light for the deep survey
(solid) and for the average survey (dotted) are plotted in histograms, with the
power-law fit for the deep survey (solid) overplotted for comparison. }

\end{figure}


\begin{figure}
\plottwo{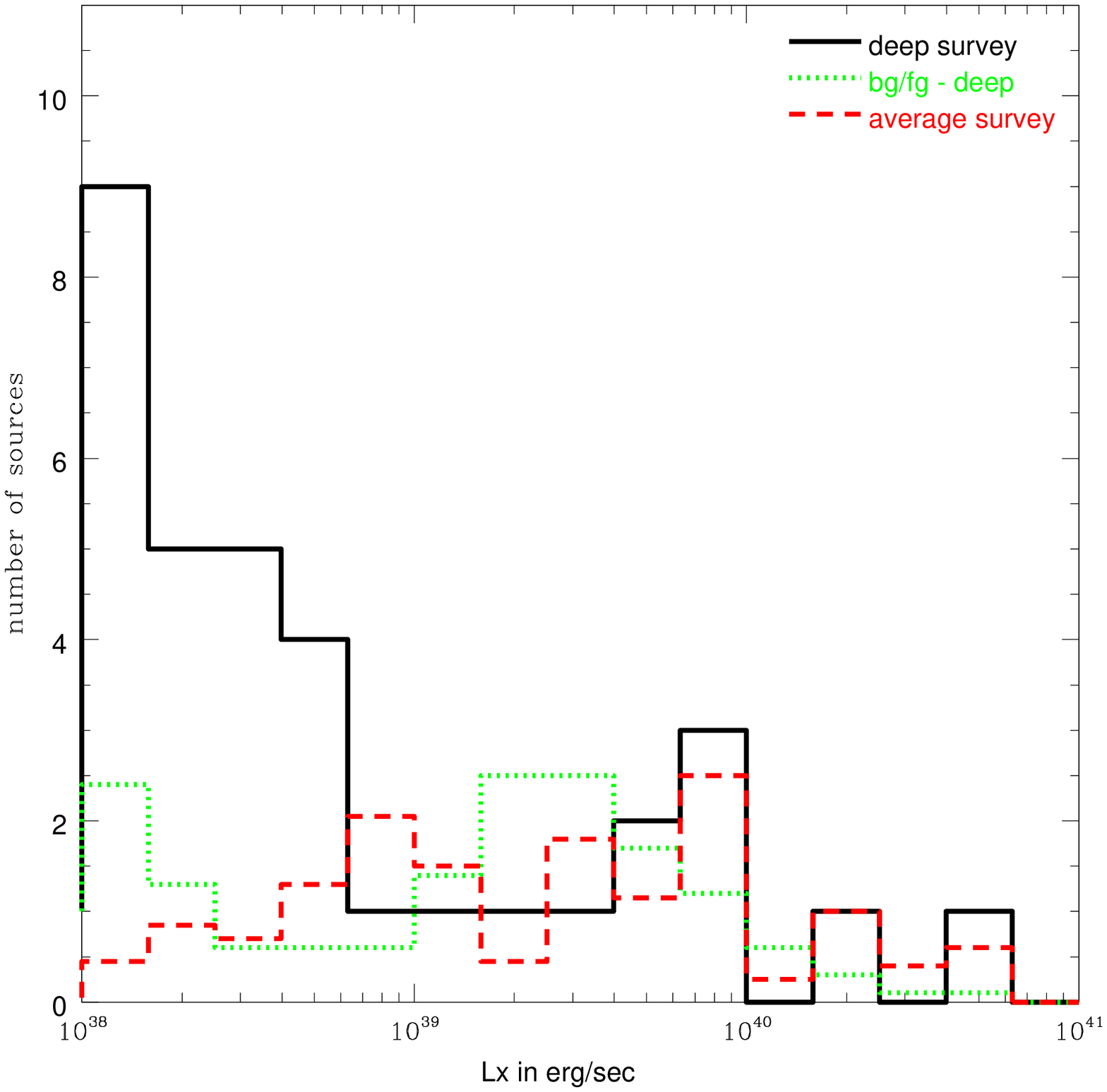}{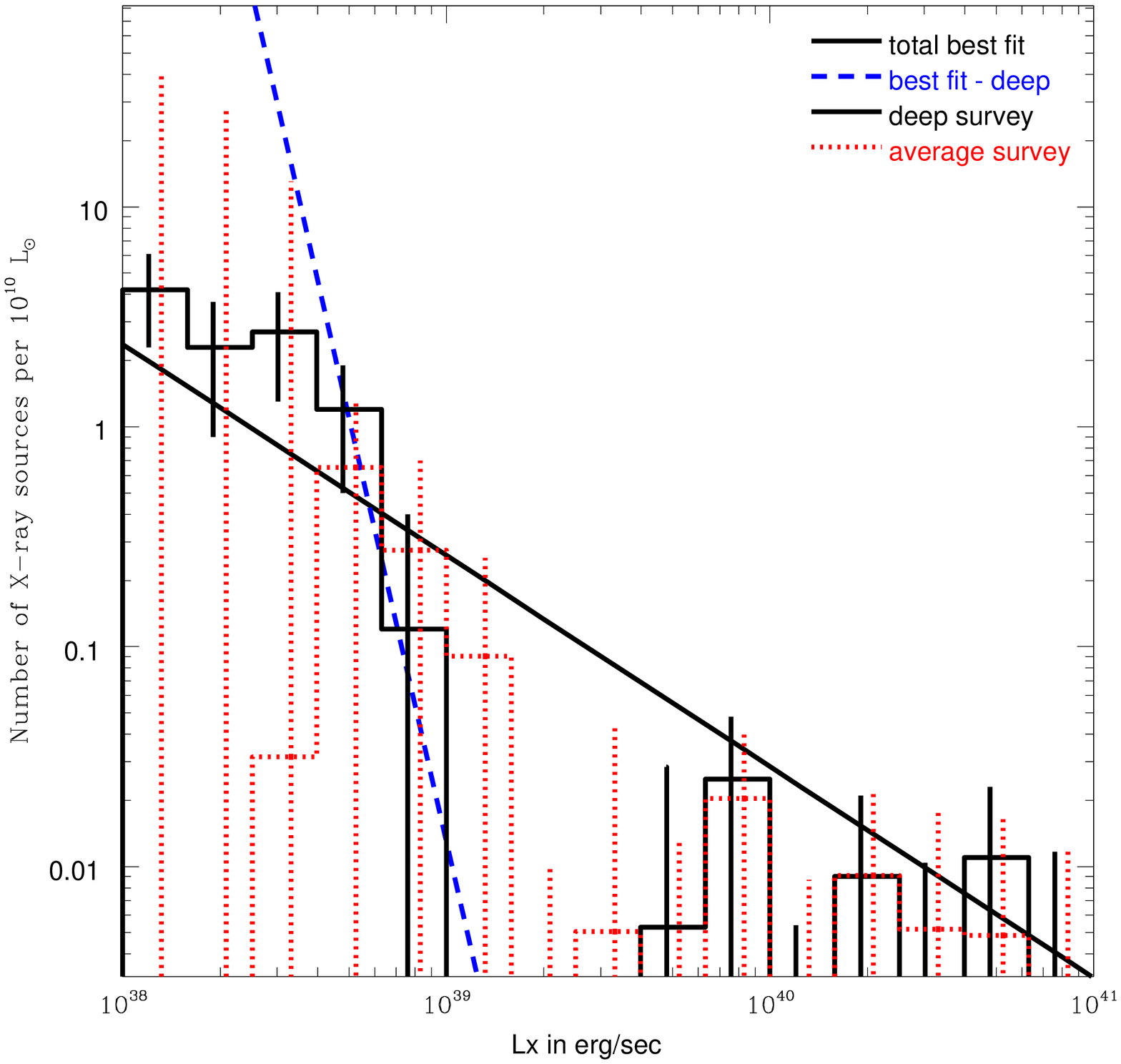}

\caption{The luminosity functions for the X-ray source population in the
early-type galaxies in the deep survey and in the average survey. (left)
Histograms of the numbers of detected ULXs (solid) and contaminating sources
(dotted) in the deep survey, and detected ULXs in the average survey (dashed).
(right) The luminosity functions scaled by the survey blue light for the deep
survey (solid) and for the average survey (dotted) are plotted in histograms,
with the power-law fit to the total X-ray population in all galaxies (solid)
and the power-law fit to the X-ray population in the early-type galaxies (dashed)
overplotted for comparison. }

\end{figure}


\begin{figure}
\plottwo{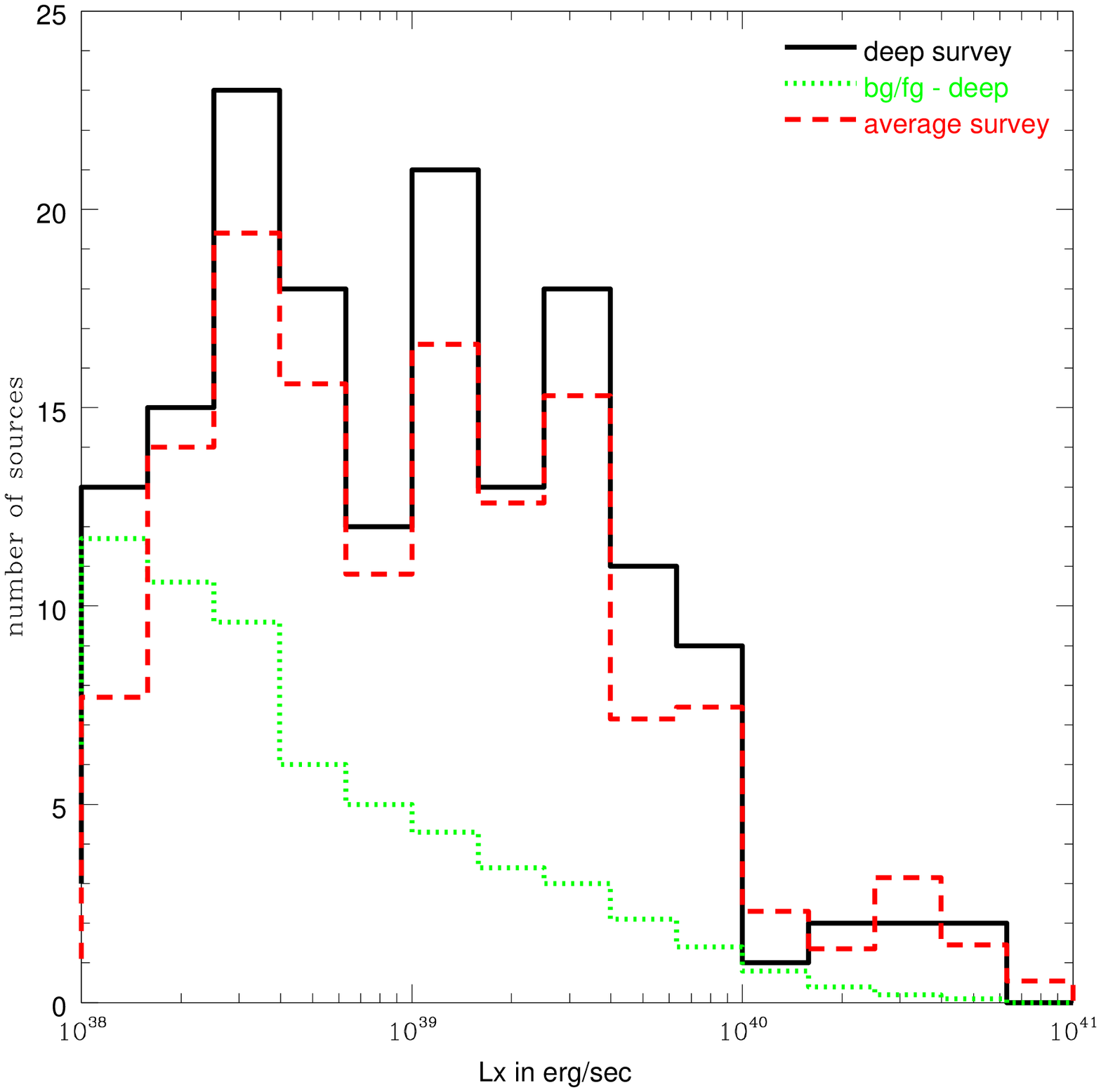}{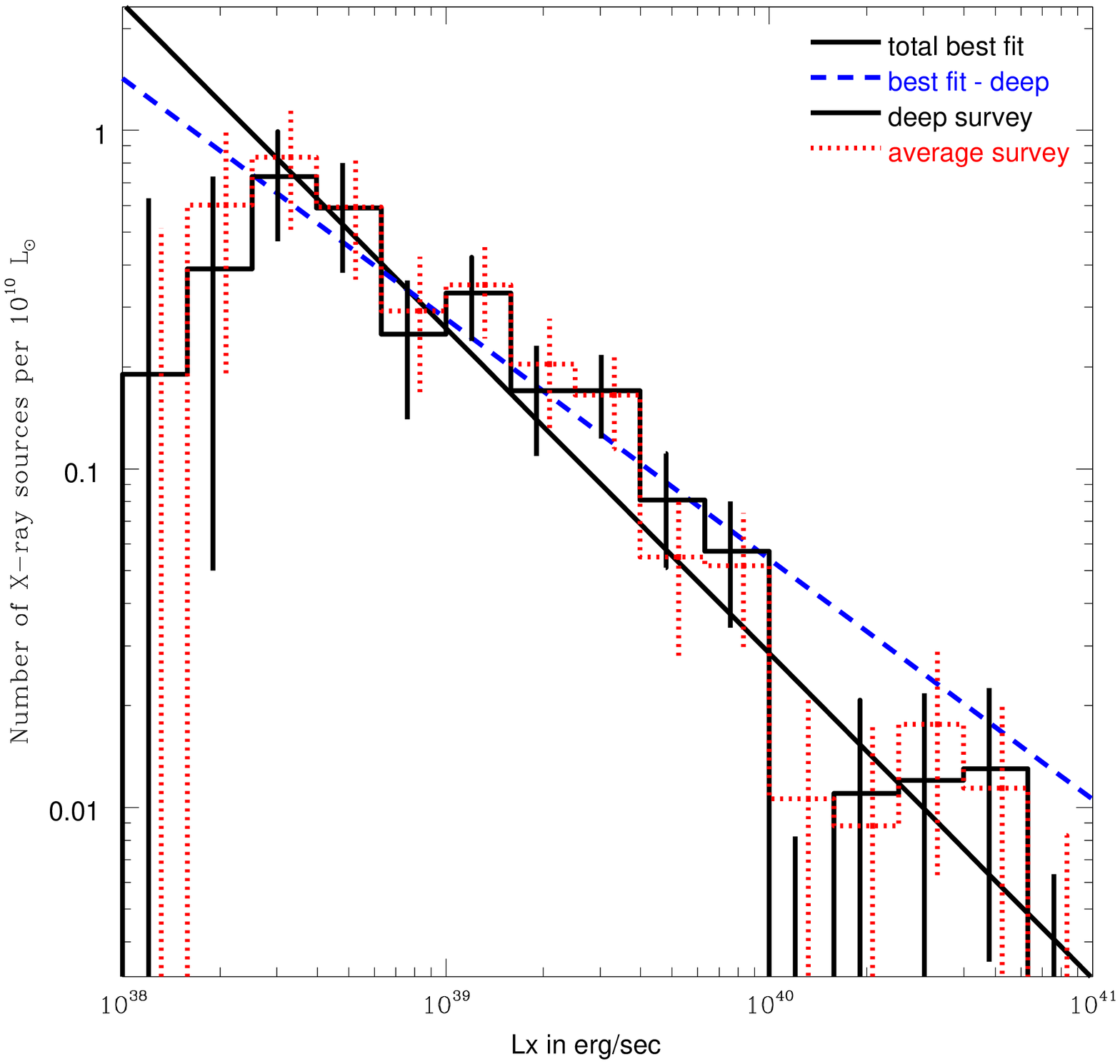}

\caption{The luminosity functions for the X-ray source populations in the
late-type galaxies in the deep survey and in the average survey. (left)
Histograms of the numbers of detected ULXs (solid) and contaminating sources
(dotted) in the deep survey, and detected ULXs in the average survey (dashed).
(right) The luminosity functions scaled by the survey blue light for the deep
survey (solid) and for the average survey (dotted) are plotted in histograms,
with the power-law fit to the total X-ray source population in all galaxies
(solid) and the power-law fit to the X-ray source population in the late-type
galaxies (dashed) overplotted for comparison. }

\end{figure}


\begin{figure}
\plotone{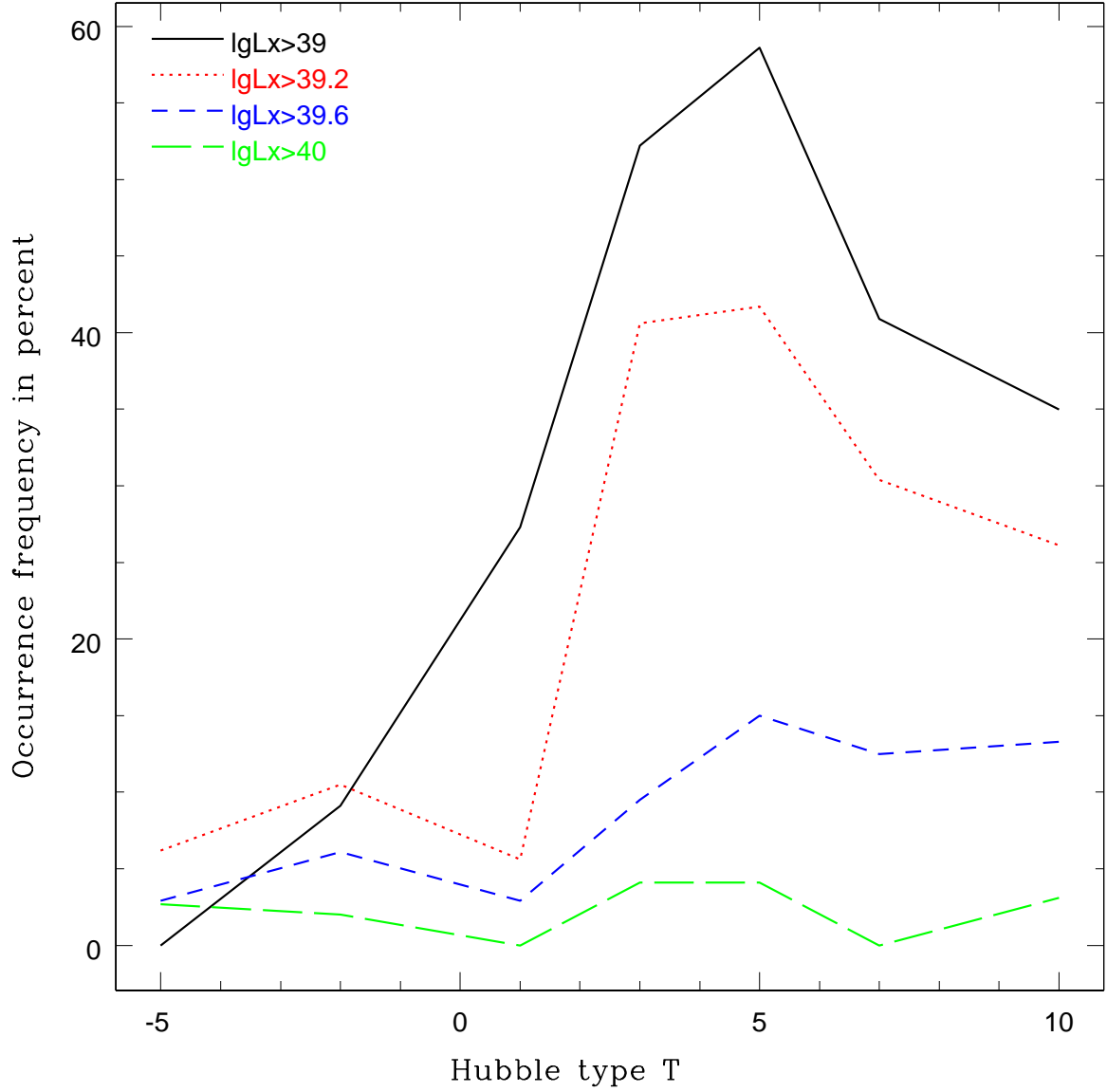}

\caption{The occurrence frequencies as a function of the Hubble type T of
galaxies for ULXs above  1/1.6/4.0/10$\times$$10^{39}$ erg/sec.}

\end{figure}


\begin{figure}
\plotone{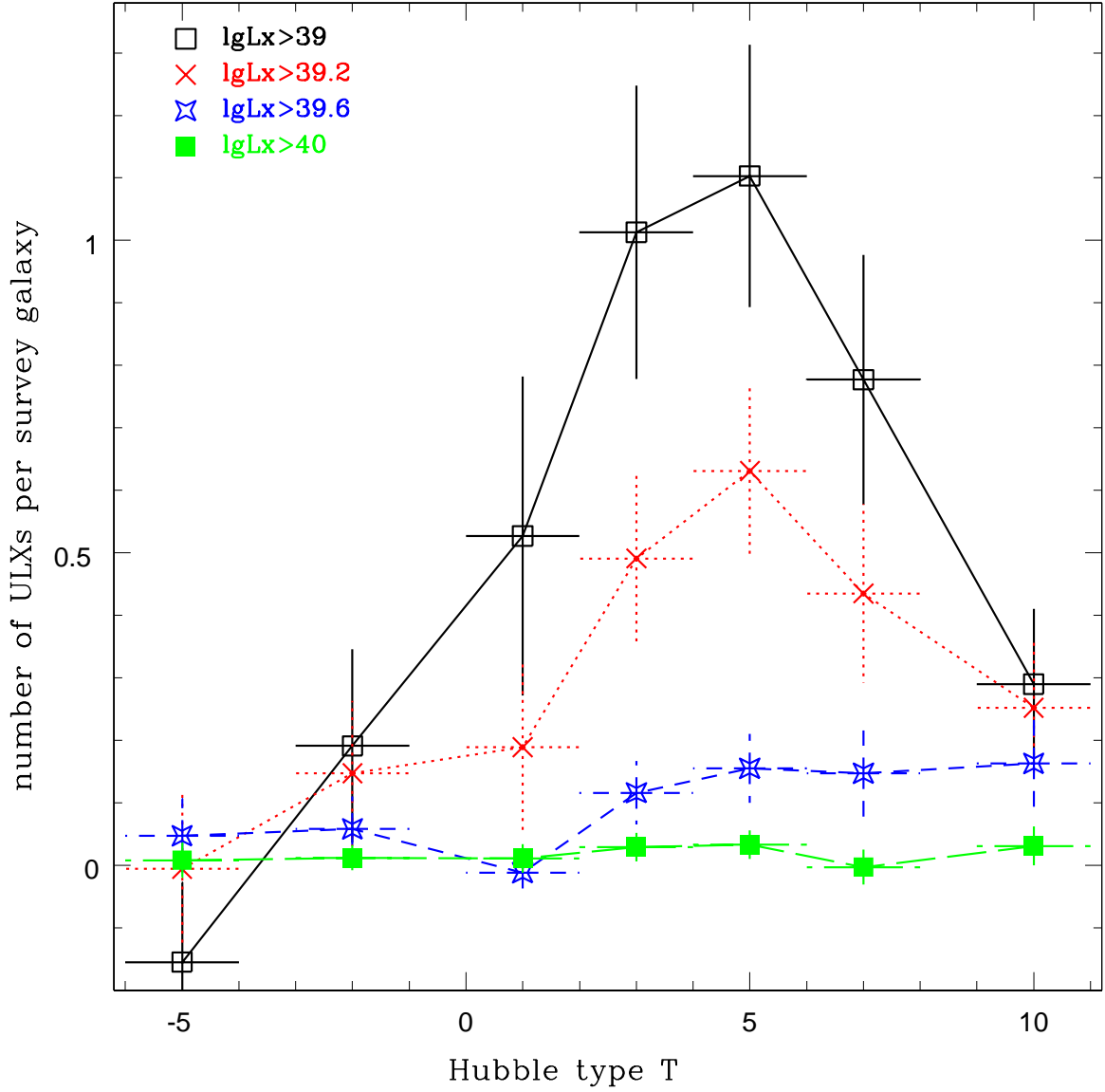}
 
\caption{The net numbers of ULXs per survey galaxy as a function of the Hubble
type T of galaxies for ULXs above 1/1.6/4.0/10$\times$$10^{39}$ erg/sec.}

\end{figure}


\begin{figure}
\plottwo{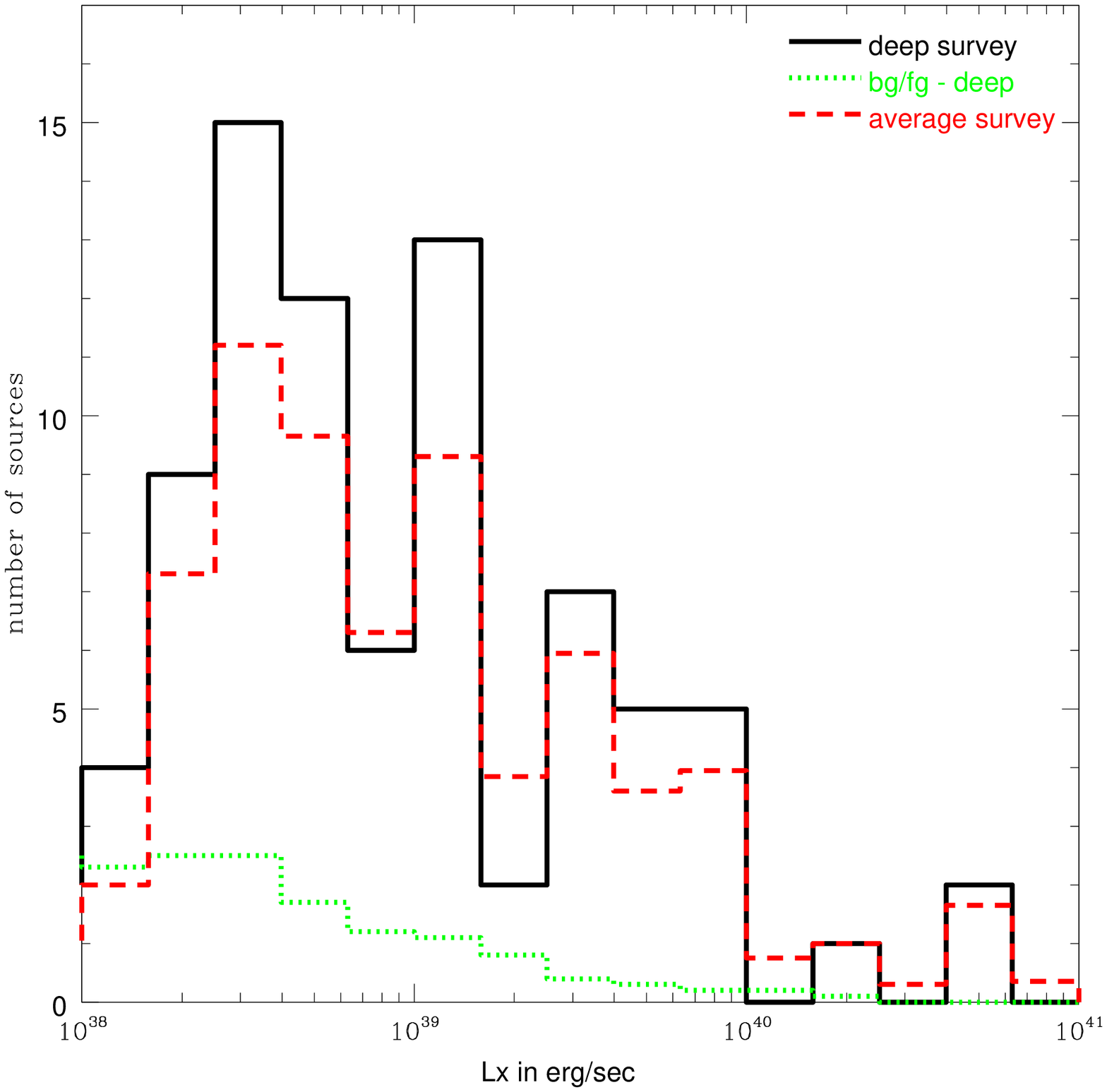}{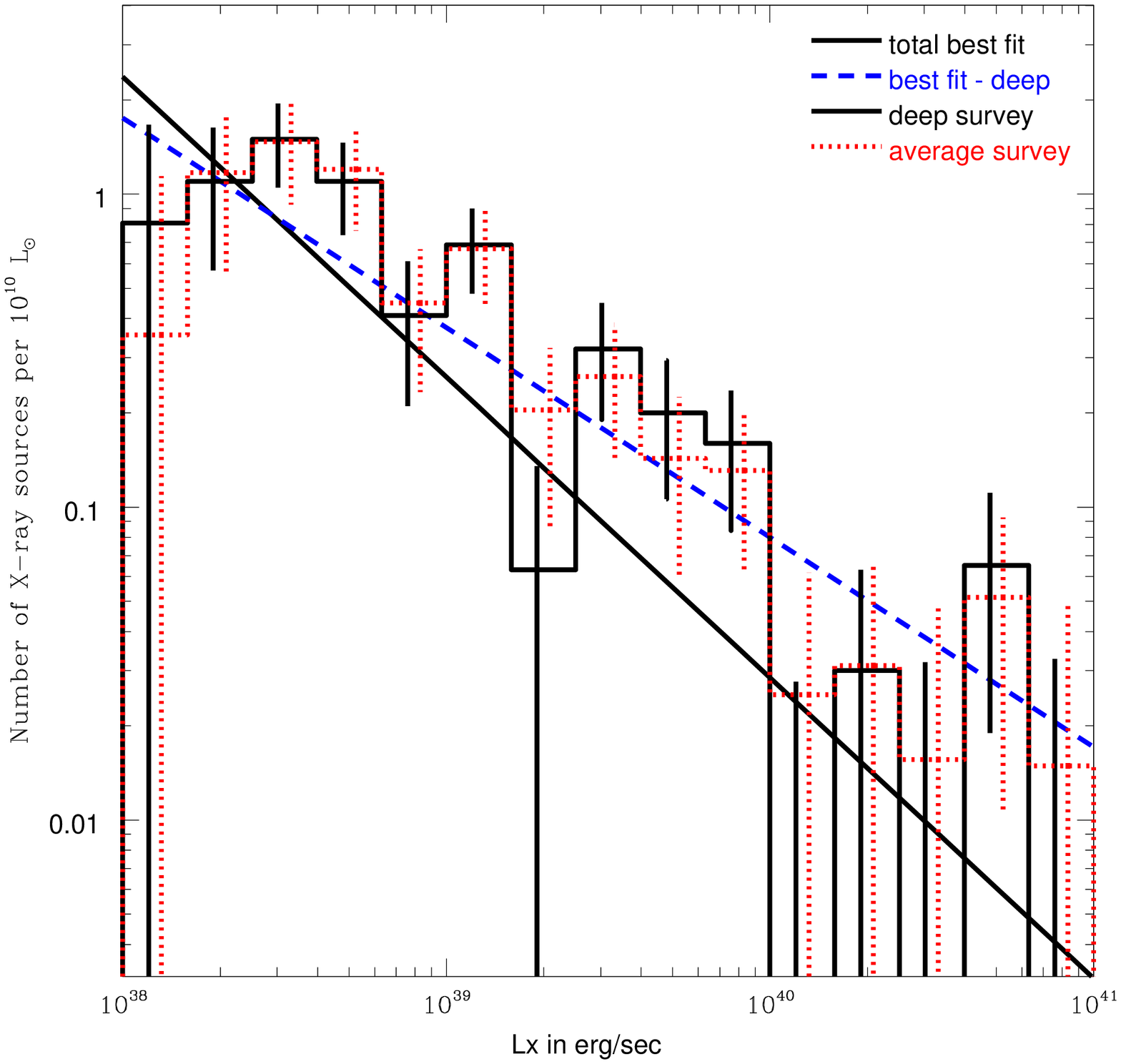}

\caption{The luminosity functions for the X-ray source populations in the
starburst/HII galaxies in the deep survey and in the average survey. (left)
Histograms of the numbers of detected ULXs (solid) and contaminating sources
(dotted) in the deep survey, and detected ULXs in the average survey (dashed).
(right) The luminosity functions scaled by the survey blue light for the deep
survey (solid) and for the average survey (dotted) are plotted in histograms,
with the power-law fit to the total X-ray population in all galaxies (solid)
and the power-law fit to the X-ray population in the starburst/HII galaxies
(dashed) overplotted for comparison. }

\end{figure}


\begin{figure}
\plottwo{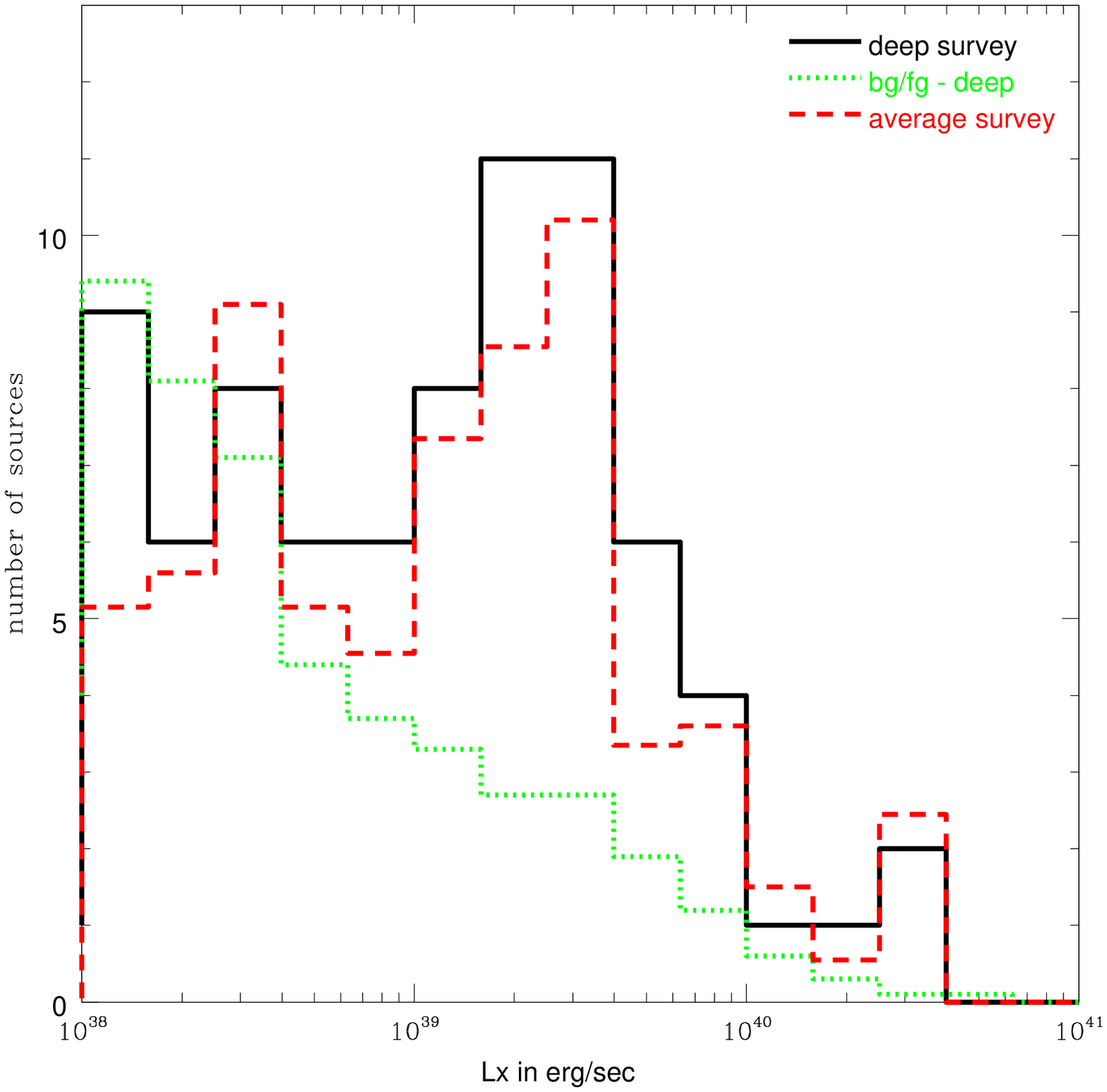}{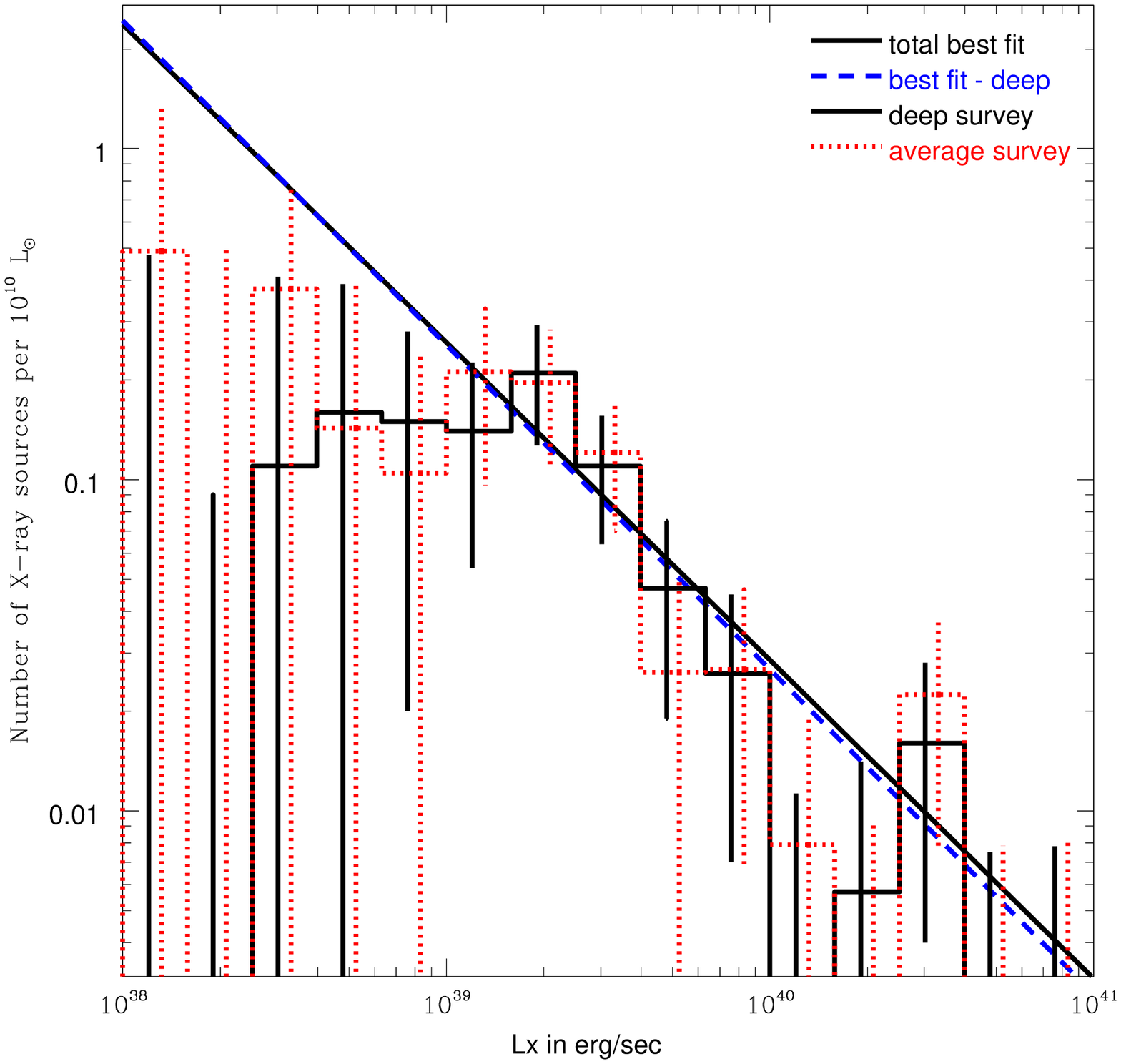}

\caption{The luminosity functions for the X-ray source populations in the
non-starburst late-type galaxies in the deep survey and in the average survey.
(left) Histograms of the numbers of detected ULXs (solid) and contaminating
sources (dotted) in the deep survey, and detected ULXs in the average survey
(dashed).  (right) The luminosity functions scaled by the survey blue light for
the deep survey (solid) and for the average survey (dotted) are plotted in
histograms, with the power-law fit to the total X-ray population in all
galaxies (solid) and the power-law fit to the X-ray population in the non-starburst
late-type galaxies (dashed) overplotted for comparison. }

\end{figure}


\begin{figure}
\plotone{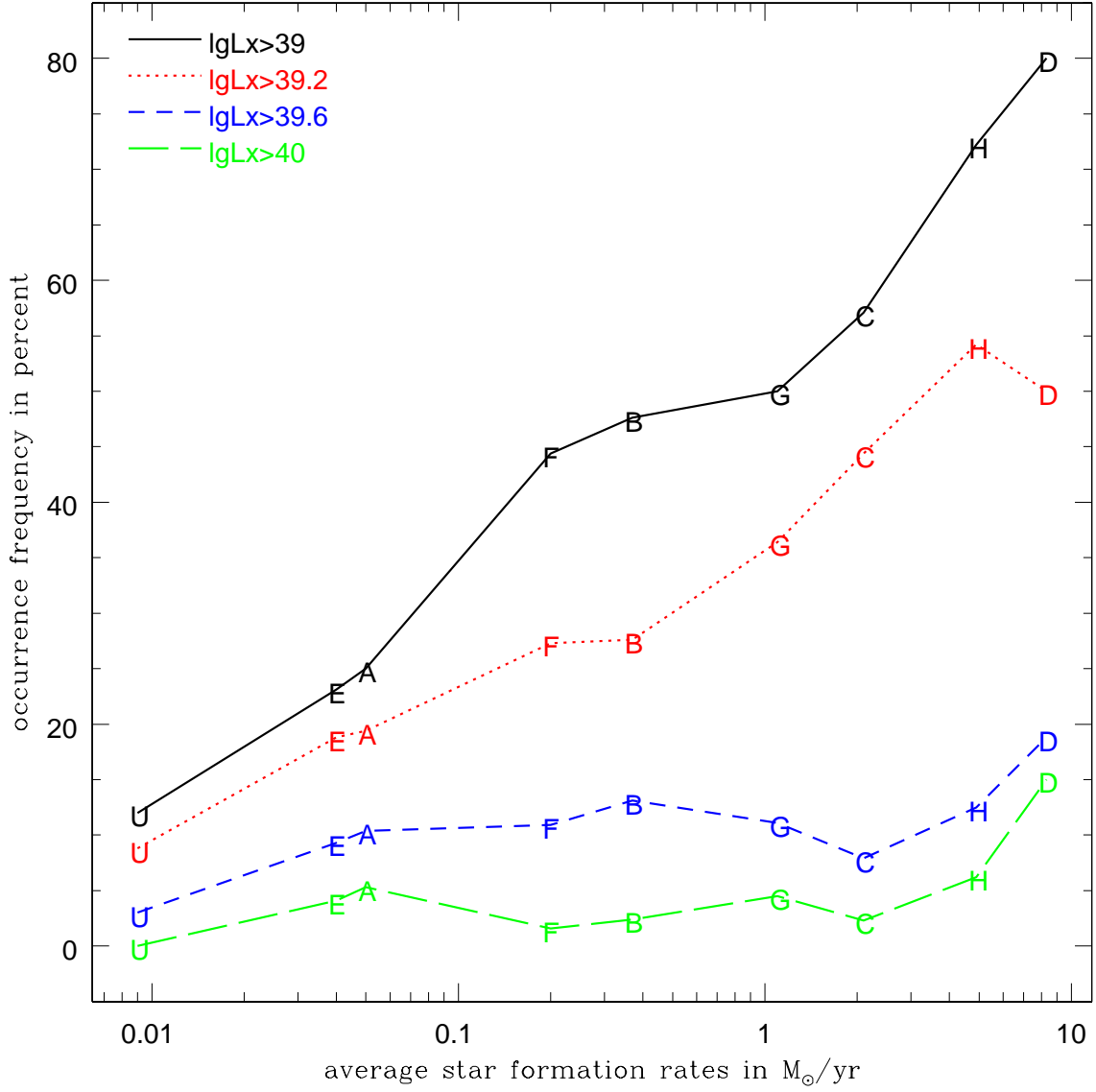}

\caption{The occurrence frequencies as a function of the star formation rates of
galaxies for ULXs above 1/1.6/4.0/10$\times$$10^{39}$ erg/sec.}

\end{figure}


\begin{figure}
\plotone{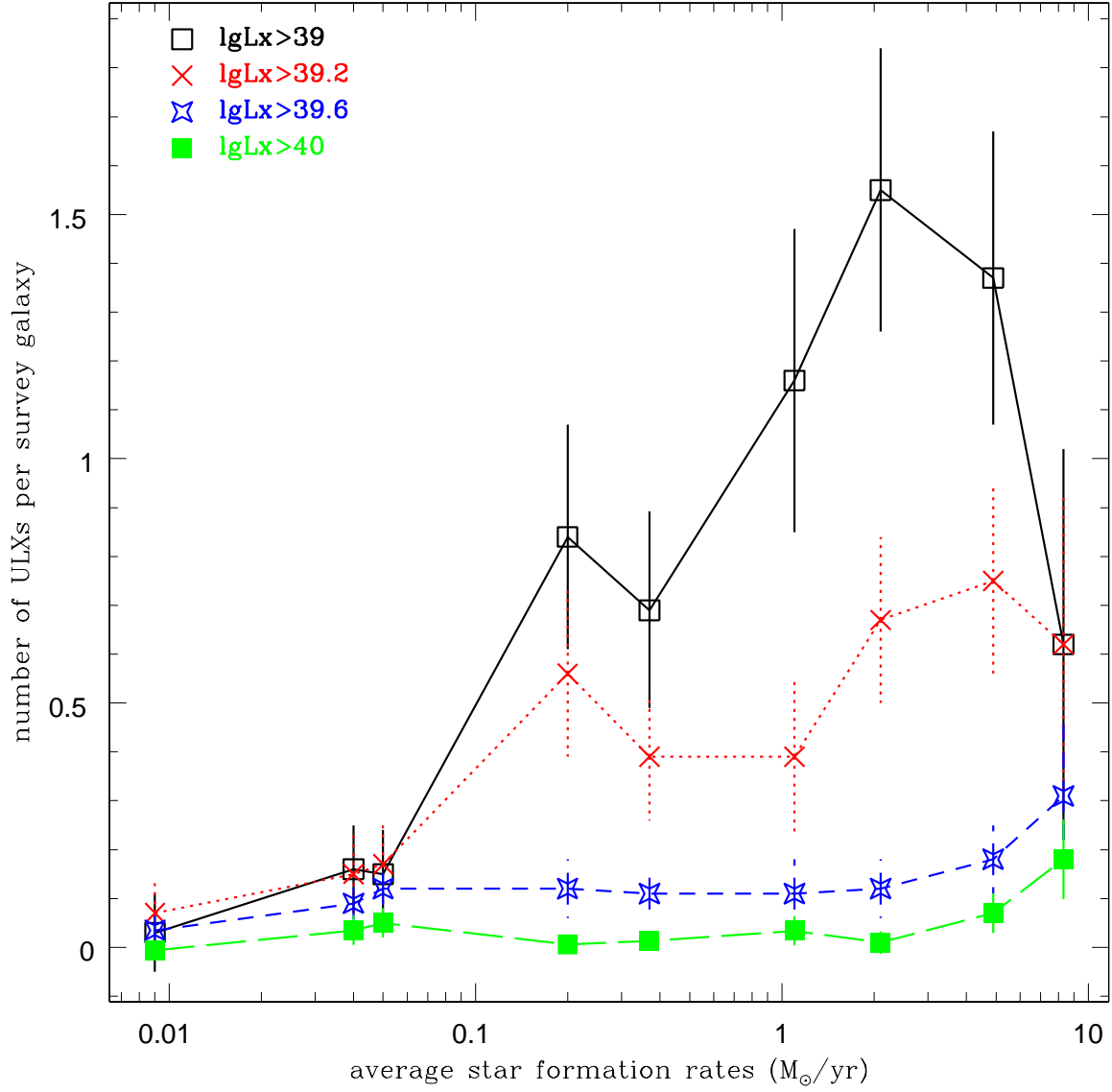}

\caption{The net numbers of ULXs per survey galaxy as a function of the star
formation rates of galaxies for ULXs above 1/1.6/4.0/10$\times$$10^{39}$
erg/sec.}

\end{figure}


\begin{figure}
\plotone{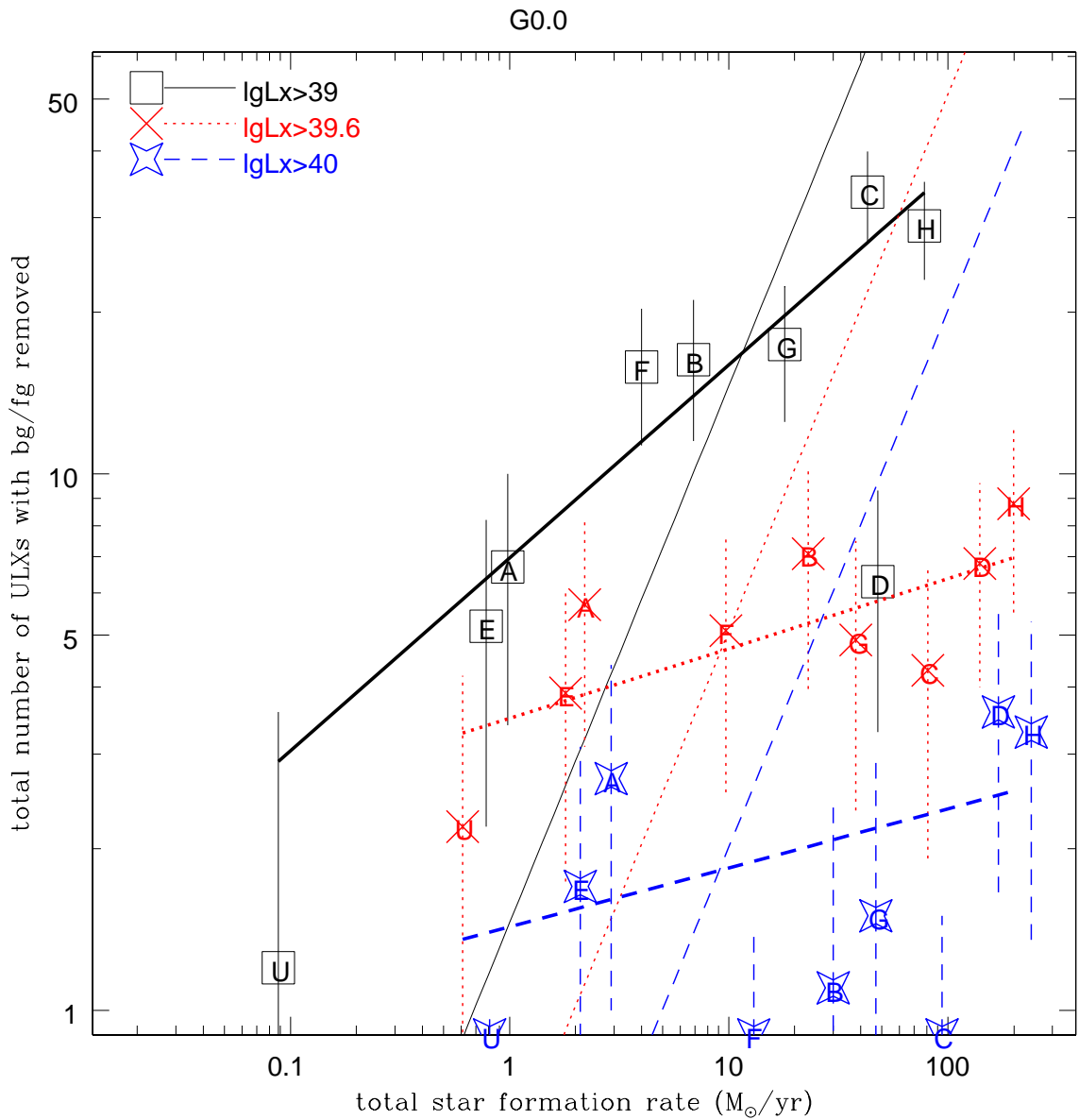}

\caption{The numbers of ULXs in a group of galaxies as a function of total star formation rates.
The fitted power-law functions are plotted in thick lines, with the $N_{ULX}$--$SFR$ relation 
from Grimm et al. (2003) overplotted in thin lines for comparison. }

\end{figure}


\begin{figure}
\plotone{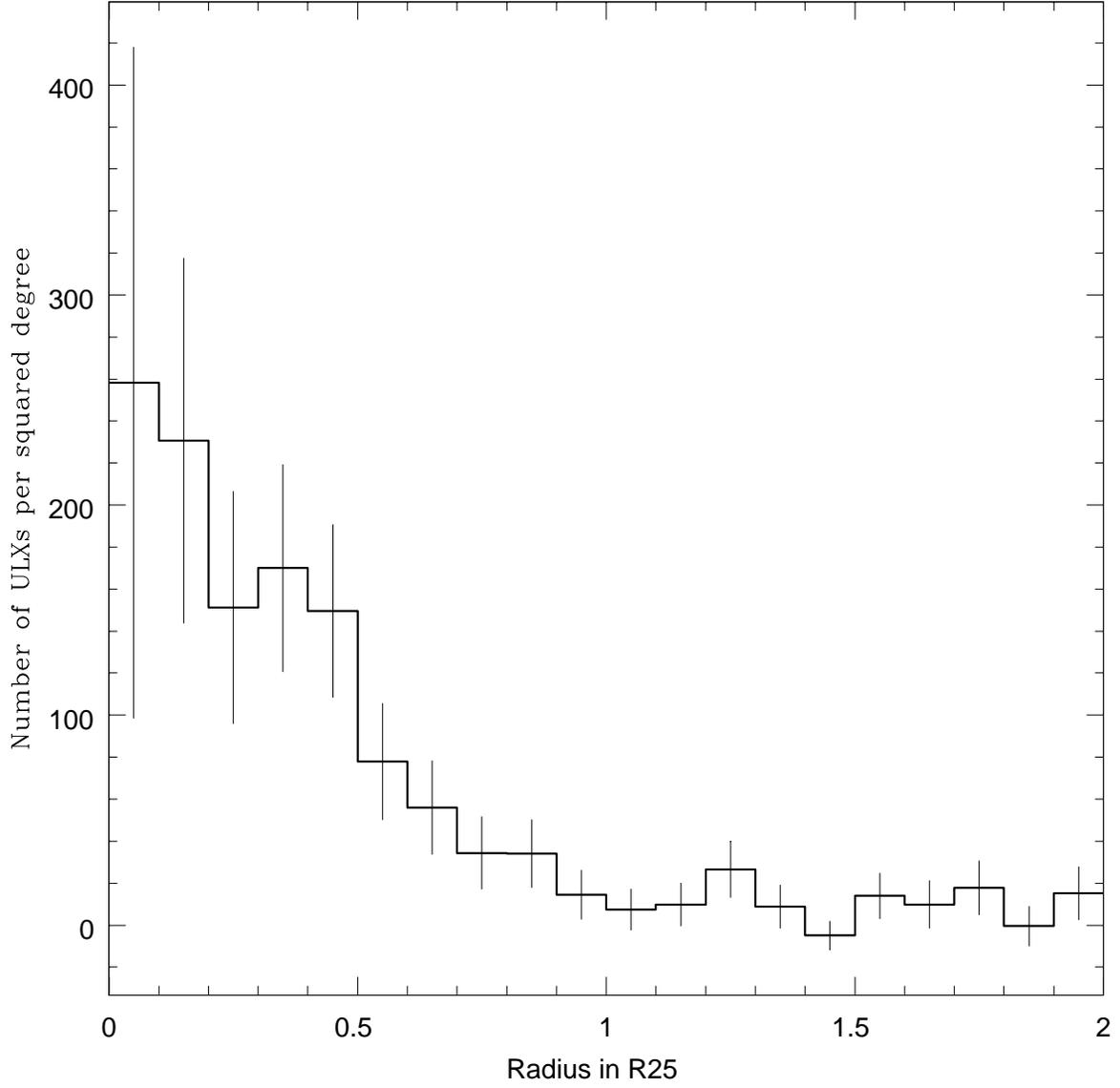}

\caption{The surface number density of ULXs in the late-type galaxies.  The
contaminating sources predicted with the $logN$--$logS$ relation are subtracted
from the detected ULXs. }

\end{figure}


\begin{figure}
\plotone{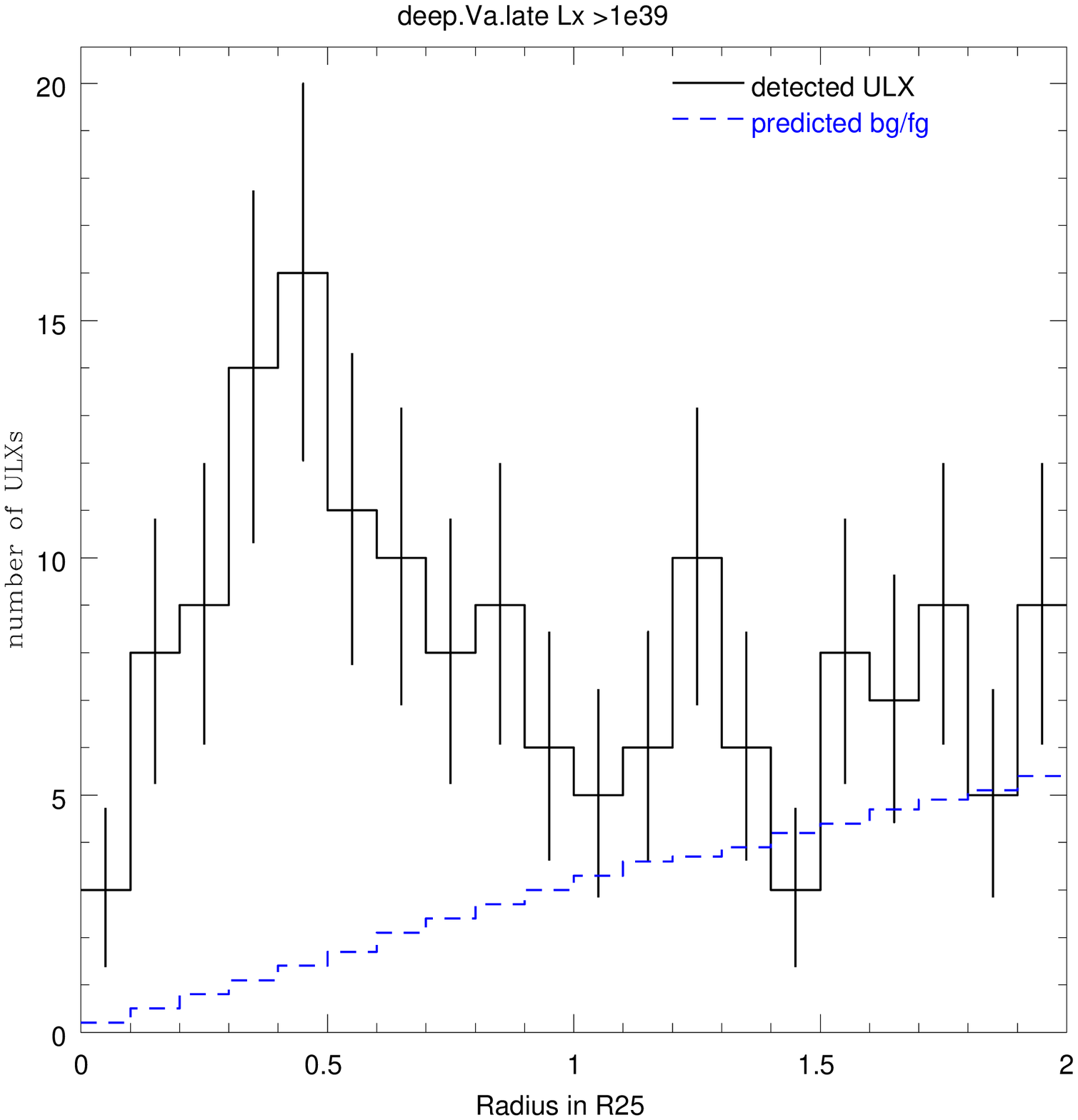}

\caption{ The radial distribution of detected ULXs and predicted contaminating
sources above $10^{39}$ erg/sec in the late-type galaxies.}

\end{figure}


\begin{figure}
\plotone{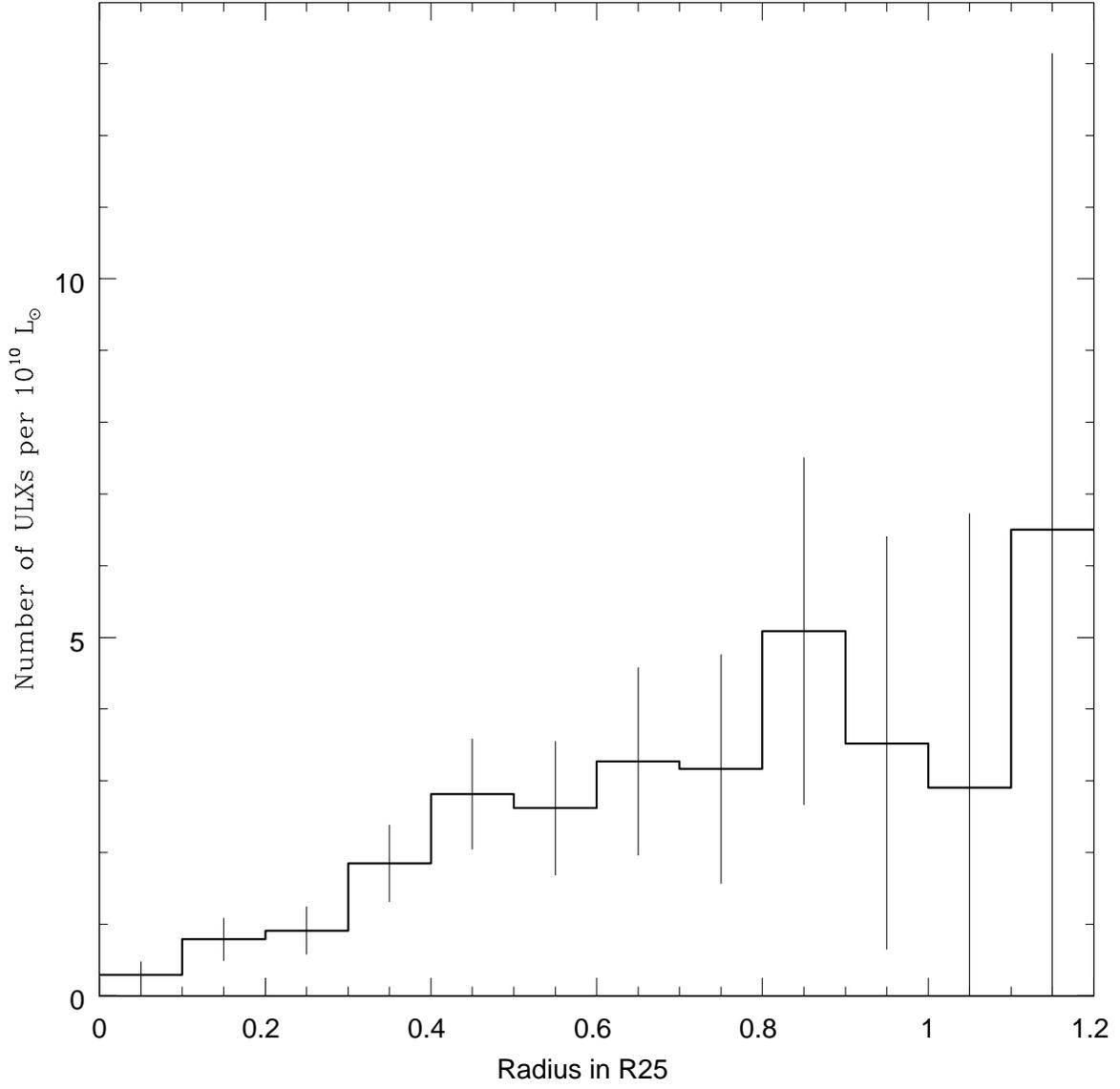}

\caption{The number of ULXs per unit blue light in the late-type galaxies.  The
contaminating sources predicted with the $logN$--$logS$ relation are subtracted
from the detected ULXs. }

\end{figure}


\begin{figure}
\plotone{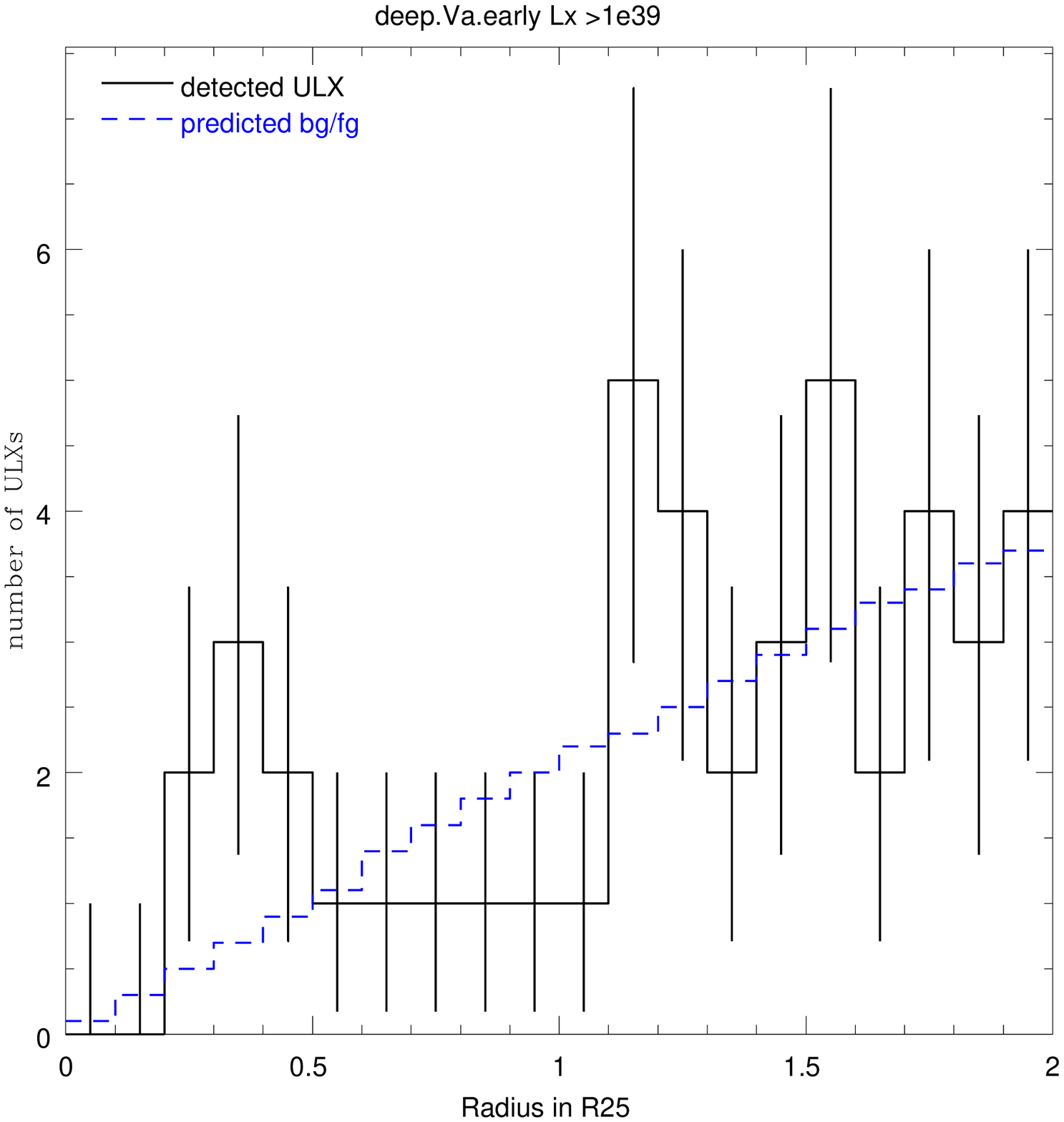}

\caption{The radial distribution of detected ULXs and predicted contaminating
sources above $10^{39}$ erg/sec in the early-type galaxies.}

\end{figure}

\clearpage

\begin{rotate}
\begin{deluxetable}{l|rrrrr|rrrrr|rrrrr|rrrrr}
\tabletypesize{\tiny}
\tablecaption{ULX occurrence rates for different survey galaxy samples}
\tablehead{
  \colhead{sample\tablenotemark{a}} &
  \multicolumn{5}{c}{$L_X \ge 10^{39}$ erg/sec\tablenotemark{b}} &
  \multicolumn{5}{c}{$L_X \ge 1.6 \times 10^{39}$ erg/sec\tablenotemark{c}} &
  \multicolumn{5}{c}{$L_X \ge 4.0 \times 10^{39}$ erg/sec\tablenotemark{d}} &
  \multicolumn{5}{c}{$L_X \ge 10^{40}$ erg/sec\tablenotemark{e}}  \\
}

\startdata

all&98&35&35.7&1.71$\pm$0.23&0.56$\pm$0.09&131&33&25.2&1.40$\pm$0.21&0.32$\pm$0.06&226&20&8.8&1.20$\pm$0.25&0.10$\pm$0.025&281&8&2.8&0.99$\pm$0.35&0.023$\pm$0.01 \\
\hline
early&22&1&4.5&1$\pm$1&0.023$\pm$0.10&35&3&8.6&1.07$\pm$0.67&0.08$\pm$0.09&67&3&4.5&1.40$\pm$0.73&0.05$\pm$0.04&88&2&2.3&1$\pm$0.70&0.01$\pm$0.016 \\
earlyx&21&1&4.8&1$\pm$1&-0.019$\pm$0.10&33&2&6.1&1.10$\pm$0.85&0.036$\pm$0.07&65&2&3.1&1.40$\pm$0.85&0.03$\pm$0.034&86&2&2.3&1$\pm$0.70&0.012$\pm$0.016 \\
late&76&34&44.7&1.73$\pm$0.23&0.72$\pm$0.10&96&30&31.2&1.43$\pm$0.22&0.41$\pm$0.07&159&17&10.7&1.16$\pm$0.27&0.12$\pm$0.03&193&6&3.1&0.98$\pm$0.40&0.03$\pm$0.013 \\
pec&3&0&0&0$\pm$0&-0.033$\pm$0.33&3&0&0&0$\pm$0&-0$\pm$0.33&4&0&0&0$\pm$0&-0.025$\pm$0.25&4&0&0&0$\pm$0&-0$\pm$0.25 \\
\hline
ellip&11&0&0&0$\pm$0&-0.155$\pm$0.13&16&1&6.2&1.20$\pm$1.40&-0.006$\pm$0.12&34&1&2.9&1.80$\pm$1.40&0.05$\pm$0.06&37&1&2.7&1$\pm$1&0.008$\pm$0.027 \\
lent&11&1&9.1&1$\pm$1&0.19$\pm$0.15&19&2&10.5&0.95$\pm$0.70&0.15$\pm$0.12&33&2&6.1&1.20$\pm$0.85&0.06$\pm$0.05&51&1&2&1$\pm$1&0.012$\pm$0.02 \\
lentx&10&1&10&1$\pm$1&0.12$\pm$0.14&17&1&5.9&1$\pm$1&0.08$\pm$0.08&31&1&3.2&1$\pm$1&0.016$\pm$0.03&49&1&2&1$\pm$1&0.016$\pm$0.02 \\
\hline
Sa&11&3&27.3&1.53$\pm$0.73&0.53$\pm$0.26&18&1&5.6&1$\pm$1&0.19$\pm$0.13&34&1&2.9&1$\pm$1&-0.012$\pm$0.03&44&0&0&0$\pm$0&0.01$\pm$0.023 \\
Sb&23&12&52.2&1.93$\pm$0.42&1.01$\pm$0.23&32&13&40.6&1.44$\pm$0.35&0.49$\pm$0.14&63&6&9.5&1.15$\pm$0.43&0.12$\pm$0.05&73&3&4.1&1$\pm$0.57&0.03$\pm$0.023 \\
Sc&29&17&58.6&2.09$\pm$0.36&1.10$\pm$0.21&36&15&41.7&1.72$\pm$0.35&0.63$\pm$0.14&60&9&15&1.20$\pm$0.37&0.15$\pm$0.06&73&3&4.1&0.97$\pm$0.57&0.03$\pm$0.023 \\
Sd&22&9&40.9&1.86$\pm$0.47&0.78$\pm$0.20&23&7&30.4&1.40$\pm$0.46&0.43$\pm$0.14&32&4&12.5&1$\pm$0.50&0.15$\pm$0.07&36&0&0&0$\pm$0&-0.003$\pm$0.03 \\
Sm&20&7&35&1.13$\pm$0.40&0.29$\pm$0.12&23&6&26.1&1.15$\pm$0.43&0.25$\pm$0.10&30&4&13.3&1.25$\pm$0.55&0.16$\pm$0.07&32&1&3.1&1$\pm$1&0.03$\pm$0.03 \\
\hline
Sbrst&26&18&69.2&1.81$\pm$0.32&0.98$\pm$0.20&29&15&51.7&1.37$\pm$0.31&0.53$\pm$0.14&40&10&25&1.19$\pm$0.35&0.26$\pm$0.08&46&2&4.3&1$\pm$0.70&0.06$\pm$0.04 \\
nSbrstL&52&16&30.8&1.65$\pm$0.34&0.56$\pm$0.12&70&15&21.4&1.51$\pm$0.33&0.34$\pm$0.08&123&7&5.7&1.11$\pm$0.40&0.07$\pm$0.03&152&4&2.6&0.97$\pm$0.50&0.018$\pm$0.013 \\
\hline
SFRU&25&3&12&1.30$\pm$0.67&0.03$\pm$0.08&34&3&8.8&1.40$\pm$0.73&0.07$\pm$0.07&66&2&3&1.90$\pm$1&0.035$\pm$0.03&83&0&0&0$\pm$0&-0.006$\pm$0.012 \\
SFRA&28&7&25&0.99$\pm$0.37&0.15$\pm$0.09&36&7&19.4&0.99$\pm$0.37&0.17$\pm$0.08&48&5&10.4&1$\pm$0.44&0.12$\pm$0.05&57&3&5.3&1$\pm$0.57&0.05$\pm$0.03 \\
SFRB&21&10&47.6&1.53$\pm$0.40&0.69$\pm$0.20&29&8&27.6&1.46$\pm$0.44&0.39$\pm$0.13&61&8&13.1&1.04$\pm$0.38&0.11$\pm$0.05&82&2&2.4&1$\pm$0.70&0.013$\pm$0.017 \\
SFRC&21&12&57.1&2.38$\pm$0.47&1.55$\pm$0.29&27&12&44.4&1.50$\pm$0.37&0.67$\pm$0.17&38&3&7.9&0.97$\pm$0.57&0.12$\pm$0.06&44&1&2.3&1$\pm$1&0.01$\pm$0.023 \\
SFRD&5&4&80&1.52$\pm$0.65&0.62$\pm$0.40&8&4&50&1.60$\pm$0.65&0.62$\pm$0.30&16&3&18.8&1.63$\pm$0.73&0.31$\pm$0.15&20&3&15&0.97$\pm$0.57&0.18$\pm$0.10 \\
SFRE&26&6&23.1&0.98$\pm$0.40&0.16$\pm$0.09&32&6&18.8&0.98$\pm$0.40&0.15$\pm$0.08&43&4&9.3&1$\pm$0.50&0.09$\pm$0.05&49&2&4.1&1$\pm$0.70&0.035$\pm$0.03 \\
SFRF&18&8&44.4&1.66$\pm$0.46&0.84$\pm$0.23&22&6&27.3&1.62$\pm$0.53&0.56$\pm$0.17&46&5&10.9&0.98$\pm$0.44&0.12$\pm$0.06&63&1&1.6&1$\pm$1&0.006$\pm$0.016 \\
SFRG&14&7&50&2.27$\pm$0.59&1.16$\pm$0.31&22&8&36.4&1.43$\pm$0.44&0.39$\pm$0.16&36&4&11.1&1.10$\pm$0.55&0.11$\pm$0.07&44&2&4.5&1$\pm$0.70&0.034$\pm$0.03 \\
SFRH&18&13&72.2&1.97$\pm$0.41&1.37$\pm$0.30&24&13&54.2&1.45$\pm$0.35&0.75$\pm$0.19&40&5&12.5&1.36$\pm$0.52&0.18$\pm$0.07&48&3&6.2&0.97$\pm$0.57&0.07$\pm$0.04 \\
\enddata

\tablenotetext{a}{The survey galaxy sample. A 'x' is suffixed when the two
peculiar lenticulars NGC1316 and NGC5128 are excluded from the early-type and
lenticular sample.}

\tablenotetext{b}{For ULXs with $L_X \ge 10^{39}$ erg/sec. The five numbers are, (1) the number of galaxies with survey light $\ge 0.1 \times$ its total light, (2) the number of galaxies with at least 0.5 net ULXs, (3) the fraction (\%) of survey galaxies that host ULXs (4) the net ULXs per host galaxy, (6) the net ULXs per survey galaxy. }

\tablenotetext{b}{For ULXs with $L_X \ge 1.6 \times 10^{39}$ erg/sec. }

\tablenotetext{b}{For ULXs with $L_X \ge 4.0 \times 10^{39}$ erg/sec. }

\tablenotetext{b}{For ULXs with $L_X \ge 10^{40}$ erg/sec. }

\end{deluxetable}

\end{rotate}

\clearpage

\begin{rotate}
\begin{deluxetable}{lrr|rrrrrrr|rrrrrrr}
\tabletypesize{\tiny}
\tablecaption{ULX rates for different survey galaxy samples}
\tablehead{
  \colhead{sample\tablenotemark{a}} & \colhead{NG\tablenotemark{b}} & \colhead{Lg\tablenotemark{c}} & \multicolumn{7}{c}{$lgL_X=$[39,39.2]} & \multicolumn{7}{c}{$lgL_X=$[39.2,39.4]} \\
  \colhead{} & \colhead{} & \colhead{} & \colhead{$ULX$\tablenotemark{d}} & \colhead{$bg$\tablenotemark{e}} & \colhead{Net\tablenotemark{f}} & \colhead{CNet\tablenotemark{g}} & \colhead{SL/SFR\tablenotemark{h}} & \colhead{Urate\tablenotemark{i}} & \colhead{Crate\tablenotemark{j}} & \colhead{$ULX$} & \colhead{$bg$} & \colhead{Net} & \colhead{CNet}  & \colhead{SL/SFR} & \colhead{Urate} & \colhead{Crate} \\
}

\startdata

all&296&0.78&22/89&5.7/26&16.3$\pm$4.7&63$\pm$9.4&59/95&0.26$\pm$0.08&0.59$\pm$0.10&14/67&5.9/20.3&8.1$\pm$3.7&46.7$\pm$8.2&91/130&0.1$\pm$0.045&0.33$\pm$0.06 \\
\hline
early&91&0.9&1/10&1.4/10.3&-0.4$\pm$1&-0.3$\pm$3.2&9.5/2.2&-0.079$\pm$0.07&-0.13$\pm$0.09&1/9&2.5/8.9&-1.5$\pm$1&0.1$\pm$3&28/3.6&-0.062$\pm$0.026&-0.047$\pm$0.05 \\
earlyx&89&0.8&1/7&1.4/8.8&-0.4$\pm$1&-1.8$\pm$2.6&8.0/0.5&-0.1$\pm$0.08&-0.19$\pm$0.09&0/6&2.2/7.4&-2.2$\pm$1&-1.4$\pm$2.4&23/1.3&-0.092$\pm$0.04&-0.087$\pm$0.05 \\
late&205&0.74&21/79&4.3/15.7&16.7$\pm$4.6&63.3$\pm$8.9&49/93&0.33$\pm$0.09&0.84$\pm$0.13&13/58&3.4/11.4&9.6$\pm$3.6&46.6$\pm$7.6&63/120&0.17$\pm$0.06&0.51$\pm$0.09 \\
pec&5&0.52&0/0&0/0.1&-0$\pm$1&-0.1$\pm$1&1.3/15&-0.03$\pm$0.75&-0.076$\pm$0.39&0/0&0/0.1&-0$\pm$1&-0.1$\pm$1&1.3/15&-0.014$\pm$0.75&-0.046$\pm$0.39 \\
\hline
ellip&38&1.4&0/4&1.1/7.2&-1.1$\pm$1&-3.2$\pm$2&5.5/0.1&-0.22$\pm$0.25&-0.34$\pm$0.04&0/4&1.9/6.1&-1.9$\pm$1&-2.1$\pm$2&20/0.4&-0.094$\pm$0.04&-0.12$\pm$0.04 \\
lent&53&0.52&1/6&0.3/3.1&0.7$\pm$1&2.9$\pm$2.4&4.4/2.1&0.1$\pm$0.17&0.19$\pm$0.21&1/5&0.6/2.8&0.4$\pm$1&2.2$\pm$2.2&8.8/3.2&0.000$\pm$0.07&0.09$\pm$0.12 \\
lentx&51&0.33&1/3&0.3/1.6&0.7$\pm$1&1.4$\pm$1.7&2.9/0.4&0.13$\pm$0.24&0.23$\pm$0.33&0/2&0.3/1.3&-0.3$\pm$1&0.7$\pm$1.4&4.3/0.9&-0.073$\pm$0.24&0.10$\pm$0.22 \\
\hline
Sa&45&0.8&2/9&1/4.8&1$\pm$1.4&4.2$\pm$3&10/4.4&0.10$\pm$0.13&0.28$\pm$0.19&3/7&1.1/3.8&1.9$\pm$1.7&3.2$\pm$2.6&16/18&0.15$\pm$0.12&0.19$\pm$0.14 \\
Sb&77&1.2&9/35&2.7/9.8&6.3$\pm$3&25.2$\pm$5.9&27/50&0.22$\pm$0.1&0.56$\pm$0.14&6/26&2/7.1&4$\pm$2.4&18.9$\pm$5.1&35/68&0.12$\pm$0.07&0.34$\pm$0.10 \\
Sc&75&0.94&13/43&2.3/7.1&10.7$\pm$3.6&35.9$\pm$6.6&24/36&0.42$\pm$0.14&1$\pm$0.2&5/30&1.6/4.7&3.4$\pm$2.2&25.3$\pm$5.5&30/53&0.12$\pm$0.08&0.59$\pm$0.13 \\
Sd&40&0.35&8/20&0.9/2.1&7.1$\pm$2.8&17.9$\pm$4.5&8.9/18&0.78$\pm$0.32&1.8$\pm$0.45&2/12&0.5/1.2&1.5$\pm$1.4&10.8$\pm$3.5&9.9/18&0.15$\pm$0.14&1$\pm$0.33 \\
Sm&36&0.077&2/9&0.1/0.4&1.9$\pm$1.4&8.6$\pm$3&1.2/0.4&1.4$\pm$1.1&4.8$\pm$1.8&2/7&0.1/0.3&1.9$\pm$1.4&6.7$\pm$2.6&1.4/0.5&1.4$\pm$1&3.4$\pm$1.4 \\
\hline
Sbrst&46&0.66&13/35&1.1/3.1&11.9$\pm$3.6&31.9$\pm$5.9&17/45&0.69$\pm$0.21&1.5$\pm$0.29&2/22&0.8/2&1.2$\pm$1.4&20$\pm$4.7&19/48&0.06$\pm$0.07&0.83$\pm$0.2 \\
nSbrstL&164&0.75&8/44&3.3/13.1&4.7$\pm$2.8&30.9$\pm$6.6&32/48&0.14$\pm$0.09&0.57$\pm$0.13&11/36&2.7/9.8&8.3$\pm$3.3&26.2$\pm$6&44/77&0.21$\pm$0.08&0.42$\pm$0.1 \\
\hline
SFRU&88&0.37&1/6&0.9/4.8&0.1$\pm$1&1.2$\pm$2.4&5.1/0.09&-0.084$\pm$0.11&-0.083$\pm$0.16&1/5&1.2/3.9&-0.2$\pm$1&1.1$\pm$2.2&11/0.14&-0.013$\pm$0.09&0.000$\pm$0.11 \\
SFRA&60&0.57&0/11&0.8/4.3&-0.8$\pm$1&6.7$\pm$3.3&7.0/1.0&-0.11$\pm$0.12&0.16$\pm$0.14&1/11&1.2/3.5&-0.2$\pm$1&7.5$\pm$3.3&16/1.5&0.01$\pm$0.08&0.27$\pm$0.14 \\
SFRB&88&0.79&5/23&1/6.7&4$\pm$2.2&16.3$\pm$4.8&14/6.9&0.28$\pm$0.16&0.65$\pm$0.21&5/18&1.3/5.7&3.7$\pm$2.2&12.3$\pm$4.2&22/11&0.21$\pm$0.12&0.37$\pm$0.14 \\
SFRC&45&1.4&16/41&2.6/7.5&13.4$\pm$4&33.5$\pm$6.4&27/43&0.47$\pm$0.14&0.96$\pm$0.19&6/25&1.8/4.8&4.2$\pm$2.4&20.2$\pm$5&33/56&0.13$\pm$0.07&0.49$\pm$0.12 \\
SFRD&20&1.7&0/9&0.3/2.7&-0.3$\pm$1&6.3$\pm$3&5.9/48&-0.057$\pm$0.18&0.25$\pm$0.18&1/9&0.4/2.4&0.6$\pm$1&6.6$\pm$3&9.7/64&0.10$\pm$0.14&0.3$\pm$0.18 \\
SFRE&51&0.56&0/9&0.8/3.8&-0.8$\pm$1&5.2$\pm$3&6.8/0.8&-0.12$\pm$0.13&0.14$\pm$0.15&1/9&1/3&-0$\pm$1&6$\pm$3&14/1.1&0.01$\pm$0.08&0.26$\pm$0.15 \\
SFRF&68&0.69&5/20&0.8/4.2&4.2$\pm$2.2&15.8$\pm$4.5&10/4&0.38$\pm$0.21&0.99$\pm$0.29&5/15&0.8/3.4&4.2$\pm$2.2&11.6$\pm$3.9&14/4.6&0.34$\pm$0.18&0.6$\pm$0.21 \\
SFRG&46&1.1&8/24&1.3/6.6&6.7$\pm$2.8&17.4$\pm$4.9&13/18&0.49$\pm$0.2&0.8$\pm$0.24&4/16&1.5/5.3&2.5$\pm$2&10.7$\pm$4&24/25&0.13$\pm$0.1&0.31$\pm$0.13 \\
SFRH&49&1.5&10/36&2/7&8$\pm$3.2&29$\pm$6&24/78&0.3$\pm$0.12&0.77$\pm$0.17&4/26&1.4/4.9&2.6$\pm$2&21.1$\pm$5.1&29/100&0.10$\pm$0.07&0.47$\pm$0.12 \\

\enddata

\tablenotetext{a}{The survey galaxy sample.  A 'x' is suffixed when the two
peculiar lenticulars NGC1316 and NGC5128 are excluded from the early-type and
lenticular sample.  }

\tablenotetext{b}{Number of galaxies in this sample.}

\tablenotetext{c}{The average blue light per galaxy in $10^{10} L_\odot$.}

\tablenotetext{d}{number of extra-nuclear X-ray sources observed in this bin, and the cumulative number.}

\tablenotetext{e}{number of predicted background/foreground X-ray sources that would have luminosity in this bin, and the cumulative number}

\tablenotetext{f}{the net number of X-ray sources in this bin, with errors as $\sqrt{N_{ULX}}$.}
 
\tablenotetext{g}{the cumulative net number of X-ray sources with luminosity of this bin and larger.}

\tablenotetext{h}{surveyed blue light in unit of $10^{10} L_\odot$  and the total star formation rates in $M_\odot/yr$for the luminosity in this bin.}

\tablenotetext{i}{the ULX rate in unit of ULX per $10^{10} L_\odot$ in this bin. }

\tablenotetext{j}{the ULX rate (ULX per $10^{10} L_\odot$) for luminoisities in
this bin and larger. }

\end{deluxetable}

\end{rotate}

\clearpage

\begin{rotate}
\begin{deluxetable}{lrr|rrrrrrr|rrrrrrr}
\tabletypesize{\tiny}
\tablecaption{ULX rates for different survey galaxy samples}
\tablehead{
  \colhead{sample\tablenotemark{a}} & \colhead{NG\tablenotemark{b}} & \colhead{Lg\tablenotemark{c}} & \multicolumn{7}{c}{$lgL_X=$[39.6,39.8]} & \multicolumn{7}{c}{$lgL_X=$[40,40.2]} \\
  \colhead{} & \colhead{} & \colhead{} & \colhead{$ULX$\tablenotemark{d}} & \colhead{$bg$\tablenotemark{e}} & \colhead{Net\tablenotemark{f}} & \colhead{CNet\tablenotemark{g}} & \colhead{SL/SFR\tablenotemark{h}} & \colhead{Urate\tablenotemark{i}} & \colhead{Crate\tablenotemark{j}} & \colhead{$ULX$} & \colhead{$bg$} & \colhead{Net} & \colhead{CNet}  & \colhead{SL/SFR} & \colhead{Urate} & \colhead{Crate} \\
}

\startdata

all&296&0.78&13/34&3.7/9&9.3$\pm$3.6&25$\pm$5.8&170/250&0.05$\pm$0.02&0.13$\pm$0.03&1/9&1.4/2.6&-0.4$\pm$1&6.4$\pm$3&219/290&-0.002$\pm$0.004&0.03$\pm$0.013 \\
\hline
early&91&0.9&2/7&1.7/3.9&0.3$\pm$1.4&3.1$\pm$2.6&60/6&0.005$\pm$0.023&0.04$\pm$0.04&0/2&0.6/1.1&-0.6$\pm$1&0.9$\pm$1.4&79/17&-0.008$\pm$0.013&0.01$\pm$0.017 \\
earlyx&89&0.8&1/5&1.3/3.4&-0.3$\pm$1&1.6$\pm$2.2&49/3.6&-0.008$\pm$0.018&0.02$\pm$0.035&0/2&0.5/1&-0.5$\pm$1&1$\pm$1.4&69/15&-0.008$\pm$0.014&0.014$\pm$0.02 \\
late&205&0.74&11/27&2.1/5&8.9$\pm$3.3&22$\pm$5.2&110/240&0.08$\pm$0.03&0.17$\pm$0.04&1/7&0.8/1.5&0.2$\pm$1&5.5$\pm$2.6&139/270&0.001$\pm$0.007&0.04$\pm$0.018 \\
pec&5&0.52&0/0&0/0.1&-0$\pm$1&-0.1$\pm$1&1.7/26&-0.007$\pm$0.75&-0.026$\pm$0.39&0/0&0/0&-0$\pm$1&-0$\pm$1&2.6/26&-0.005$\pm$0.39&-0.009$\pm$0.39 \\
\hline
ellip&38&1.4&1/4&1.1/2.7&-0.1$\pm$1&1.3$\pm$2&40/1.4&-0.005$\pm$0.02&0.023$\pm$0.04&0/1&0.4/0.8&-0.4$\pm$1&0.2$\pm$1&53/1.7&-0.008$\pm$0.02&0.005$\pm$0.02 \\
lent&53&0.52&1/3&0.6/1.3&0.4$\pm$1&1.7$\pm$1.7&19/4.5&0.024$\pm$0.05&0.07$\pm$0.08&0/1&0.2/0.4&-0.2$\pm$1&0.6$\pm$1&26/16&-0.007$\pm$0.04&0.023$\pm$0.04 \\
lentx&51&0.33&0/1&0.2/0.7&-0.2$\pm$1&0.3$\pm$1&9.2/2.2&-0.026$\pm$0.1&0.004$\pm$0.06&0/1&0.1/0.2&-0.1$\pm$1&0.8$\pm$1&15/13&-0.008$\pm$0.06&0.045$\pm$0.06 \\
\hline
Sa&45&0.8&1/2&0.7/1.6&0.3$\pm$1&0.4$\pm$1.4&28/56&0.007$\pm$0.03&0.008$\pm$0.04&0/1&0.3/0.5&-0.3$\pm$1&0.5$\pm$1&35/68&-0.007$\pm$0.03&0.015$\pm$0.03 \\
Sb&77&1.2&5/11&1.3/3.1&3.7$\pm$2.2&7.9$\pm$3.3&69/150&0.05$\pm$0.03&0.1$\pm$0.04&0/3&0.5/0.9&-0.5$\pm$1&2.1$\pm$1.7&84/160&-0.005$\pm$0.012&0.023$\pm$0.02 \\
Sc&75&0.94&7/13&0.8/2&6.2$\pm$2.6&11$\pm$3.6&49/87&0.13$\pm$0.05&0.2$\pm$0.07&1/3&0.3/0.6&0.7$\pm$1&2.4$\pm$1.7&65/110&0.01$\pm$0.014&0.034$\pm$0.025 \\
Sd&40&0.35&3/5&0.2/0.4&2.8$\pm$1.7&4.6$\pm$2.2&11/19&0.24$\pm$0.15&0.39$\pm$0.19&0/0&0/0.1&-0$\pm$1&-0.1$\pm$1&12/20&-0.004$\pm$0.08&-0.008$\pm$0.07 \\
Sm&36&0.077&1/5&0.1/0.1&0.9$\pm$1&4.9$\pm$2.2&2.3/0.7&0.42$\pm$0.45&2$\pm$0.93&0/1&0/0&-0$\pm$1&1$\pm$1&2.5/0.8&-0.006$\pm$0.41&0.38$\pm$0.4 \\
\hline
Sbrst&46&0.66&5/13&0.3/0.8&4.7$\pm$2.2&12.2$\pm$3.6&23/72&0.2$\pm$0.09&0.45$\pm$0.13&0/3&0.2/0.3&-0.2$\pm$1&2.7$\pm$1.7&30/92&-0.005$\pm$0.03&0.09$\pm$0.06 \\
nSbrstL&164&0.75&6/14&1.9/4.4&4.1$\pm$2.4&9.6$\pm$3.7&89/170&0.05$\pm$0.03&0.10$\pm$0.04&1/4&0.6/1.3&0.4$\pm$1&2.7$\pm$2&112/190&0.003$\pm$0.009&0.023$\pm$0.017 \\
\hline
SFRU&88&0.37&1/4&0.7/1.8&0.3$\pm$1&2.2$\pm$2&22/0.6&0.004$\pm$0.036&0.07$\pm$0.07&0/0&0.3/0.5&-0.3$\pm$1&-0.5$\pm$1&31/0.8&-0.009$\pm$0.03&-0.017$\pm$0.03 \\
SFRA&60&0.57&2/7&0.6/1.3&1.4$\pm$1.4&5.7$\pm$2.6&30/2.2&0.045$\pm$0.05&0.18$\pm$0.08&0/3&0.2/0.3&-0.2$\pm$1&2.7$\pm$1.7&33/2.9&-0.005$\pm$0.03&0.08$\pm$0.05 \\
SFRB&88&0.79&4/10&1.1/2.9&2.9$\pm$2&7.1$\pm$3.2&45/23&0.07$\pm$0.045&0.13$\pm$0.06&0/2&0.5/0.9&-0.5$\pm$1&1.1$\pm$1.4&65/30&-0.007$\pm$0.015&0.016$\pm$0.02 \\
SFRC&45&1.4&2/6&0.8/1.7&1.2$\pm$1.4&4.3$\pm$2.4&48/81&0.025$\pm$0.03&0.08$\pm$0.05&0/1&0.3/0.5&-0.3$\pm$1&0.5$\pm$1&56/94&-0.005$\pm$0.018&0.008$\pm$0.017 \\
SFRD&20&1.7&4/8&0.5/1.2&3.5$\pm$2&6.8$\pm$2.8&24/140&0.15$\pm$0.08&0.24$\pm$0.1&1/4&0.2/0.4&0.8$\pm$1&3.6$\pm$2&33/170&0.024$\pm$0.03&0.11$\pm$0.06 \\
SFRE&51&0.56&2/5&0.6/1.1&1.4$\pm$1.4&3.9$\pm$2.2&25/1.8&0.06$\pm$0.06&0.14$\pm$0.08&0/2&0.1/0.3&-0.1$\pm$1&1.7$\pm$1.4&28/2.1&-0.005$\pm$0.035&0.06$\pm$0.05 \\
SFRF&68&0.69&3/7&0.6/1.9&2.4$\pm$1.7&5.1$\pm$2.6&26/9.7&0.09$\pm$0.07&0.16$\pm$0.08&0/1&0.3/0.6&-0.3$\pm$1&0.4$\pm$1&44/13&-0.008$\pm$0.02&0.007$\pm$0.02 \\
SFRG&46&1.1&2/7&1/2.1&1$\pm$1.4&4.9$\pm$2.6&45/38&0.02$\pm$0.03&0.10$\pm$0.05&0/2&0.3/0.5&-0.3$\pm$1&1.5$\pm$1.4&50/47&-0.005$\pm$0.02&0.03$\pm$0.027 \\
SFRH&49&1.5&5/11&0.9/2.2&4.1$\pm$2.2&8.8$\pm$3.3&53/200&0.08$\pm$0.04&0.15$\pm$0.06&1/4&0.4/0.7&0.6$\pm$1&3.3$\pm$2&66/240&0.01$\pm$0.015&0.05$\pm$0.03 \\

\enddata

\tablenotetext{a}{The survey galaxy sample.  A 'x' is suffixed when the two
peculiar lenticulars NGC1316 and NGC5128 are excluded from the early-type and
lenticular sample.  }

\tablenotetext{b}{Number of galaxies in this sample.}

\tablenotetext{c}{The average blue light per galaxy in $10^{10} L_\odot$.}

\tablenotetext{d}{number of extra-nuclear X-ray sources observed in this bin, and the cumulative number.}

\tablenotetext{e}{number of predicted background/foreground X-ray sources that would have luminosity in this bin, and the cumulative number}

\tablenotetext{f}{the net number of X-ray sources in this bin, with errors as $\sqrt{N_{ULX}}$.}
 
\tablenotetext{g}{the cumulative net number of X-ray sources with luminosity of this bin and larger.}

\tablenotetext{h}{surveyed blue light in unit of $10^{10} L_\odot$ and the total star formation rate in $M_\odot/yr$ for the luminosity in this bin.}

\tablenotetext{i}{the ULX rate in unit of ULX per $10^{10} L_\odot$ in this bin. }

\tablenotetext{j}{the ULX rate (ULX per $10^{10} L_\odot$) for luminoisities in
this bin and larger. }

\end{deluxetable}

\end{rotate}

\begin{deluxetable}{llccrr}
\tabletypesize{\small}
\tablecaption{Power-law fits to luminosity functions}
\tablehead{
 \colhead{galaxies} &\colhead{$lgL_1$--$lgL_2$\tablenotemark{a}} &\colhead{$\alpha$} &\colhead{$\beta$} &\colhead{$\chi^2_\nu$\tablenotemark{b}} &\colhead{$Q_\nu$\tablenotemark{c}}
}
\startdata
all &38.4--40 &0.56 +0.19 -0.19 &1.96 +0.40 -0.29 &0.85 &0.53\\
early &38.4--40 &0.07 +0.08 -0.07 &4.81 +0.63 -2.46 &1.20 &0.3\\
late &38.4--40 &0.62 +0.21 -0.21 &1.73 +0.36 -0.25 &0.76 &0.6\\
Sbrst\tablenotemark{d} &38.6--40 &1.08 +0.47 -0.47 &1.73 +0.68 -0.35 &0.70 &0.62\\
nSbrstL\tablenotemark{e} &39--40 &0.55 +0.24 -0.24 &1.98 +0.49 -0.27 &0.63 &0.59\\
\enddata
\tablenotetext{a}{The luminosity interval in which the fit is carried out.}

\tablenotetext{b}{The reduced $\chi^2$. For a good fit, this is a number not much larger or smaller than one.}

\tablenotetext{c}{The goodness of fit. The model is usually not acceptable if $Q_\nu \ll 1$.}

\tablenotetext{d}{The fit is for the luminosity function in the average survey
of starburst galaxies, while for all the first three groups of galaxies the
luminosity functions in the deep surveys are used.}

\tablenotetext{e}{The fit is for the luminosity function in the average survey
of non-starburst late-type galaxies. }

\end{deluxetable}

\begin{deluxetable}{lcccl}
\tabletypesize{\small}
\tablecaption{Grouping galaxies based on different properties}
\tablehead{
 \colhead{Group} &\colhead{No. of gal.} &\colhead{lower limit} &\colhead{upper limit} &\colhead{average} 
}
\startdata
\multicolumn{5}{c}{group galaxies based on the Hubble type T} \\
\hline
early & 91 & -6 & -1 &  \\
late  & 205 & 0 & 11 &  \\
peculiar & 5 & 90 & 99 & \\
elliptical &  38 & -6 & -4 & -5 \\
lenticular & 53 & -3 & -1 & -2 \\
Sa & 45 & 0 & 2 & 1 \\
Sb & 77 & 2 & 4 & 3 \\
Sc & 75 & 4 & 6 & 5 \\
Sd & 40 & 6 & 8 & 7 \\
Sm & 36 & 9 & 11 & 10 \\
\hline
\multicolumn{5}{c}{group galaxies based on star formation rates ($M_\odot/yr$)} \\
\hline
SFRU\tablenotemark{a} & 102 & \nodata & \nodata & $<0.009$ \\
SFRA & 60 & \nodata & $0.1$ & $0.05$ \\
SFRB & 88 & $0.1$ & $1$ & $0.4$ \\
SFRC & 46 & $1$ & $4$ & $2.1$ \\
SFRD & 21 & $4$ & \nodata & $8.3$ \\
SFRE & 51 & \nodata & $0.09$ & $0.04$ \\
SFRF & 68 & $0.09$ & $0.5$ & $0.2$ \\
SFRG & 47 & $0.5$ & $1.8$ & $1.1$ \\
SFRH & 50 & $1.8$ & \nodata & $4.9$ \\

\enddata
\tablenotetext{a}{This group includes galaxies with only upper limits in the flux 
densities at $60 \mu$. }

\end{deluxetable}


\begin{thebibliography}{}

\bibitem{} Angelini, L., Loewenstein, M., and Mushotzky, R.F., 2001, ApJL, 557, 35

\bibitem{} Bauer, F.E., Brandt, W.N., Sambruna, R.M., Chartas, G., Garmire,
G.P., Kaspi, S., and Netzer, H., 2001, AJ, 122, 182

\bibitem{} Begelman, M.C., 2002, ApJ, 568, L97

\bibitem{} Bell, E., and de Jong, R., 2001, ApJ, 550, 212

\bibitem{} Colbert, E. J. M. and Mushotzky, R. F. 1999, ApJ, 519, 89

\bibitem{} Colbert, E. and Ptak, A. 2002, ApJS, 143,25

\bibitem{} Colbert, E., Heckman, T., Ptak, A., and Strickland, D., 2004, ApJ, 602, 231

\bibitem{} de Vaucouleurs, G., de Vaucouleurs, A., Corwin, H., Buta, R., Paturel, G., and  Fouque, P, 1991, {\it Third Reference Catalogue of Bright Galaxies}

\bibitem{} Fabbiano, G. 1989, ARA\&A, 27, 87

\bibitem{} Gao, Y., Wang, Q., Appleton, P., and Lucas, R., 2003, ApJL, 596, 171

\bibitem{} Graham, A., 2001, AJ, 121, 820

\bibitem{} Grimm, H., Gilfanov, M., and Sunyaev, R., 2003, MNRAS, 339, 793

\bibitem{} Hasinger, G., Burg, R., Giacconi, R., Schmidt, M., Trumper, J. and Zamorani, G. . 1998, A\&A, 329, 482

\bibitem{} Irwin, J., Bregman, J., and Athey, A., 2004, ApJL, 60, 143

\bibitem{} Jordan, A, Cote, P., Ferrarese, L., Blakeslee, J., et al. 2004, ApJ, 613, 279

\bibitem{} Liu, J., and Bregman, J., 2005, ApJS, in press

\bibitem{} Liu, J., Bregman, J., Lloyd-Davies, E., Irwin, J., Espaillat, C., and Seitzer, P., 2004,  submitted

\bibitem{} Liu, J., Bregman, J., and Seitzer, P., 2002, ApJL, 580, 31

\bibitem{} Liu, J., Bregman, J., and Irwin, J., 2002, ApJL, 581, 93

\bibitem{} Kilgard, R., Kaaret, P., Krauss, M., Prestwich, A., Raley, M., and Zezas, A., 2002, ApJ, 573, 138

\bibitem{} King, A.R., Davies, M.B., Ward, M.J., Fabbiano, G., and Elvis, M.
2001, ApJ, 552, L109

\bibitem{} King, A.R., 2002, MNRAS, 335, 513

\bibitem{} Kording, E., Falcke, H., Markoff, S., 2002, A\&A, 382, L13

\bibitem{} Makishima, K., Kubota, A., Muzuno, T., Ohnishi, T., Tashiro, M., et
al. 2000, ApJ, 535, 632

\bibitem{} Miller, J., Fabbiano, G., Miller, M., and Fabian, A., 2003, ApJ, 585, 37

\bibitem{} Miller, J., Fabian, A., and Miller, M., 2004, ApJ, 607, 931

\bibitem{} Pakull, M.W. and Mirioni, L., 2002, astro-ph/0202488

\bibitem{} Ptak, A., and Colbert, E., 2004, ApJ, 606, 291

\bibitem{} Rice, W., Lonsdale, C, Soifer, B., Neugebauer, G., Koplan, E., Lloyd, L., de Jong, T., and Habing, H., 1988, ApJS, 68, 91

\bibitem{} Roberts, T. P. and Warwick, R. S. 2000, MNRAS, 315, 98 (RW2000)

\bibitem{} Rosa-González D., , Terlevich, E., and Terlevich, R., 2002, MNRAS, 332, 283

\bibitem{} Shakura, N., and Sunyaev, R., 1973, A\&A, 24, 337

\bibitem{} Soria, R., and Motch, C., 2004, A\&A, 422, 915

\bibitem{} Soria, R., Motch, C., Read, A., and Stevens, I, 2004, A\&A, 423, 955
 
\bibitem{} Strohmayer, T., and Mushotzky, R., 2003, ApJL, 586, 61L

\bibitem{} Sugiho, M., Kotoku, J., Makishima, K., Kubota, A., Mizuno, T.,
Fukazawa, Y., and Tashiro, M. 2001, ApJ, 561, L73

\bibitem{} Swartz, D., Ghosh, K., Tennant, A., and Wu, K., 2004, ApJS, 154, 519

\bibitem{} Wu, H., Xue, S., Xia, X., Deng, Z., and Mao, S., 2002, ApJ, 576, 738

\bibitem{} Zezas, A., Fabbiano, G., Rots, A., and Murray, S., 2002, ApJS, 142, 239

\end{thebibliography}
\end{document}